\UseRawInputEncoding
\documentclass[journal]{IEEEtran}

\usepackage{amsfonts}
\usepackage{slashbox}
\usepackage{amsmath}
\usepackage{multirow}
\usepackage{graphicx}
\usepackage{amsfonts}
\usepackage{amsmath,epsfig}
\usepackage{mathbbold}
\usepackage{stfloats}
\usepackage{amssymb}
\usepackage{url}
\usepackage{bm}
\usepackage{cite}
\usepackage{amsmath}
\usepackage{amssymb}
\usepackage{latexsym}
\usepackage{multirow}
\usepackage{epsfig}
\usepackage{graphics}
\usepackage{mathrsfs}
\usepackage{algorithmic}
\usepackage{algorithm}
\usepackage{subfigure}
\usepackage{slashbox}
\usepackage[all]{xy}

\usepackage{pifont}
\usepackage{bbding}
\usepackage{stmaryrd}
\usepackage{amssymb}
\usepackage{amsfonts}
\usepackage{epic}
\usepackage{stfloats}
\usepackage{latexsym}
\usepackage{epstopdf}
\usepackage{epic}
\usepackage{bm}
\usepackage{xcolor}
\usepackage{mathrsfs}
\usepackage{pifont}
\usepackage{bbding}
\usepackage{amsmath,epsfig}
\usepackage{mathbbold}
\usepackage{stmaryrd}
\usepackage{amssymb}
\usepackage{amsfonts}
\usepackage{epic}
\usepackage{graphicx}
\usepackage{subfigure}

\usepackage{enumerate}

\usepackage{stfloats}
\usepackage{latexsym}
\usepackage{epstopdf}
\usepackage{epic}
\usepackage{multirow}
\usepackage{stfloats}
\usepackage{bm}
\usepackage{amsmath}
\usepackage{amssymb}
\usepackage{amsthm}

\usepackage{booktabs}
\usepackage{color}
\usepackage{cases}
\usepackage{array}
\usepackage{makecell}
\usepackage{times}

\DeclareMathOperator{\maxo}{\mathrm{maximize}}

\DeclareMathAlphabet\mathbfcal{OMS}{cmsy}{b}{n}

\graphicspath{{./figs/}}

\begin{document}
\title{A Comprehensive Overview on 5G-and-Beyond Networks with UAVs: From Communications to Sensing and Intelligence}
\author{\IEEEauthorblockN{Qingqing Wu, Jie Xu, Yong~Zeng,
Derrick Wing Kwan Ng,  Naofal Al-Dhahir, Robert Schober,  and A. Lee Swindlehurst‬
}
\thanks{Q. Wu is with the State Key Laboratory of Internet of Things for Smart City and Department of Electrical and Computer Engineering, University of Macau, Macau, China 999078 (email: qingqingwu@um.edu.mo). }
\thanks{J. Xu is with the Future Network of Intelligence Institute (FNii) and the School of Science and Engineering, The Chinese University of Hong Kong (Shenzhen), Shenzhen 518172, China (e-mail: xujie@cuhk.edu.cn).}
\thanks{Y. Zeng is with the National Mobile Communications Research Laboratory, Southeast University, Nanjing 210096, and also with the Purple Mountain Laboratories, Nanjing 211111, China (e-mail:
yong\_zeng@seu.edu.cn).}
\thanks{D. W. K. Ng is with the School of Electrical Engineering and Telecommunications, the University of New South Wales, NSW 2052, Australia (e-mail: w.k.ng@unsw.edu.au).}
\thanks{N. Al-Dhahir is with the Department of Electrical and Computer Engineering, the University of Texas at Dallas, TX 75083-0688, USA (e-mail: aldhahir@utdallas.edu).}
\thanks{R. Schober is with the Institute for Digital Communications, Friedrich-Alexander-University Erlangen-Nurnberg, 91058 Erlangen, Germany (e-mail: robert.schober@fau.de).}
\thanks{A. L. Swindlehurst is with the Center for Pervasive Communications and Computing, University of California, Irvine, CA 92697, USA (e-mail: swindle@uci.edu).}
\thanks{The work of Q. Wu was supported in part by the Macau Science and Technology Development Fund, Macau SAR, under SKL-IOTSC-2021-2023, 0119/2020/A3, 0108/2020/A, and the Guangdong NSF under Grant 2021A1515011900.   The work of J. Xu was supported by the National Natural Science Foundation of China under grants 61871137 and U2001208, and the Science and Technology Program of Guangdong Province under grant 2021A0505030002.  The work of Y. Zeng was supported by the Natural Science Foundation of China under Grant 62071114, by the Fundamental Research Funds for the Central Universities of China under grant number 3204002004A2, and also by the ``Program for Innovative Talents and Entrepreneur in Jiangsu" under grant number 1104000402. The work of D. W. K. Ng was supported by funding from the UNSW Digital Grid Futures Institute, UNSW, Sydney, under a cross-disciplinary fund scheme and by the Australian Research Council's Discovery Project (DP210102169).  The work of  A. L. Swindlehurst was supported by U.S. National Science Foundation grant ECCS-2030029. }
}

\maketitle

\begin{abstract}
Due to the advancements in cellular technologies and the dense deployment of cellular infrastructure, integrating  unmanned aerial vehicles (UAVs) into the fifth-generation (5G) and beyond cellular networks is a promising solution to achieve safe UAV operation as well as enabling diversified applications with mission-specific payload data delivery. In particular, 5G networks need to support three typical usage scenarios, namely, enhanced mobile broadband (eMBB), ultra-reliable low-latency communications (URLLC), and massive machine-type communications (mMTC). On the one hand, UAVs can be leveraged as cost-effective aerial platforms to provide ground users with enhanced communication services by exploiting their high cruising altitude and controllable maneuverability in  three-dimensional (3D) space. On the other hand, providing such communication services simultaneously for both UAV and ground users  poses new challenges due to the need for ubiquitous 3D signal coverage as well as the strong air-ground network interference. Besides the requirement of high-performance wireless communications,  the ability to support effective and efficient sensing as well as network intelligence is also essential for 5G-and-beyond 3D heterogeneous wireless networks with coexisting aerial and ground users.  In this paper, we provide a comprehensive overview of the latest research efforts on integrating UAVs into cellular networks, with an emphasis on how to exploit  advanced techniques (e.g., intelligent reflecting surface, short packet transmission, energy harvesting, joint communication and radar sensing, and edge intelligence) to meet the diversified service requirements of next-generation wireless systems.  Moreover, we highlight important directions for further investigation in future work.
\end{abstract}


\section{Introduction}

\begin{figure*}[!t]
\centering
\includegraphics[width=0.7\textwidth]{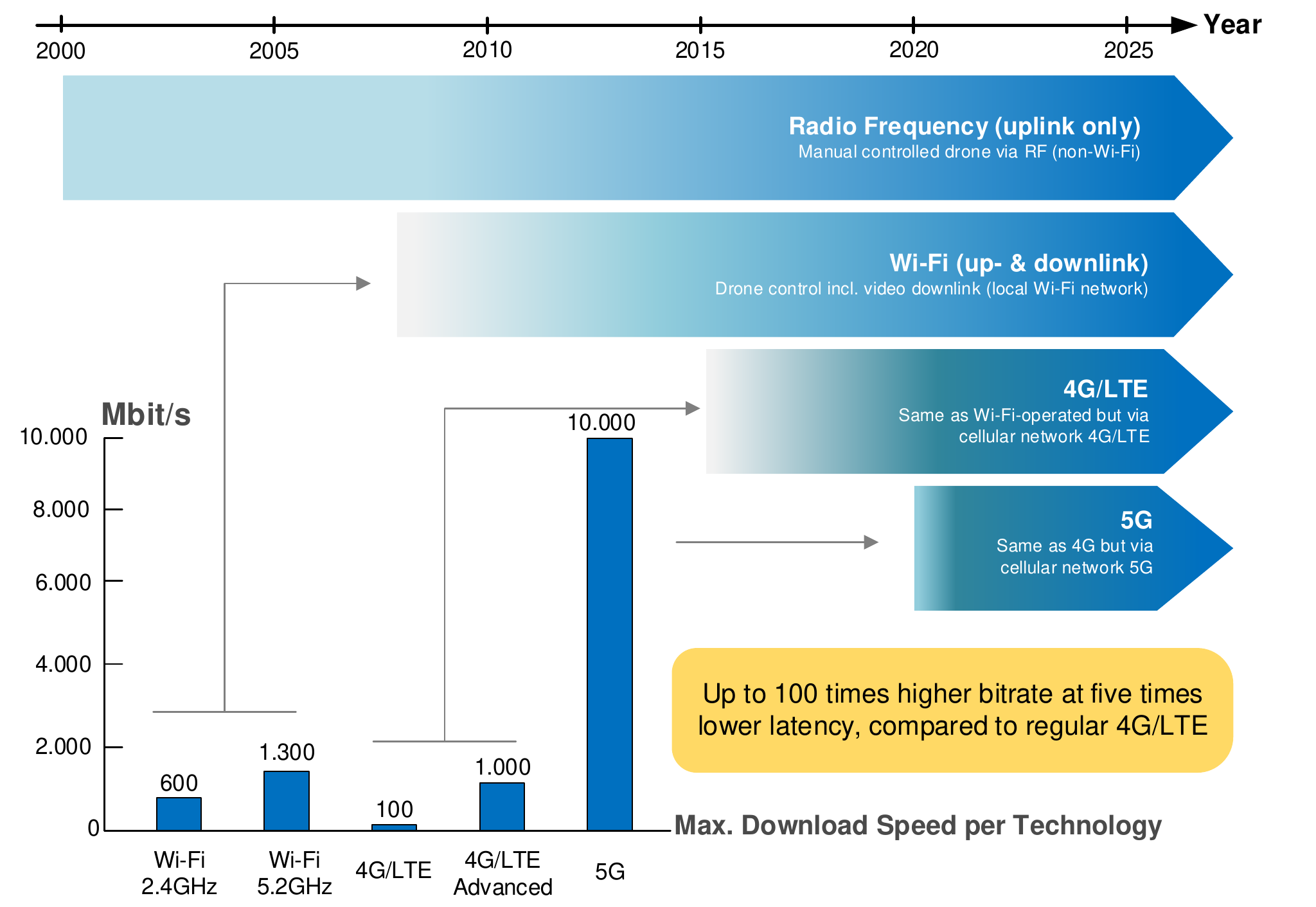}
\caption{The evolution of drone connectivity and the role of 5G \cite{drone5G}. } \label{system:model}
\end{figure*}
The global market for commercial unmanned aerial vehicles (UAVs), also known as drones,  has grown significantly over the last decade and is projected to skyrocket  to $45.8$  billion dollars in 2025 from $19.3$  billion dollars in 2020 \cite{UAVmarket2019}.  The major driving factors behind such a dramatic market size growth are  the steadily decreasing manufacturing costs and the increasing number of applications in a broad range of civilian and  commercial sectors,  including  surveillance and monitoring, aerial imaging,  precision agriculture, smart logistics, law enforcement, disaster response, and prehospital emergency care.
Particularly, as announced by the Federal Aviation Administration (FAA) during its press conference on the ``Drone Integration Pilot Program'' in Washington on November 8, 2017 \cite{USAdrone2017}: ``The need for us to integrate unmanned aircraft  into the National Airspace System (NAS) continues to be a national priority. After the hurricanes, drones became a literal lifeline. They gave us an operational window that was a game changer at every level.'' This national program was launched to further explore the expanded use of drones, including beyond-visual-line-of-sight (BVLOS) flights, night-time operation, and flights over people. Later, during the Notre Dame  Cathedral fire in 2019 \cite{NotreDameCathedralfire2019}, two UAVs equipped with high-resolution thermal imaging cameras were dispatched in BVLOS environments to help firefighters gauge the scene through the billowing smoke and effectively position firehoses to combat the blaze in real-time. Very recently, UAVs have also been applied worldwide to combat the spread of COVID-19, e.g., to  facilitate communication/broadcast information, disinfect outbreak-affected areas, deliver critical supplies, and measure body temperatures \cite{COVID19drone}. {In December 28, 2020, the FAA further released two long-awaited drone rules. One is on remote identification for UAVs and the other is on UAV operations at night and over people. These new rules are expected to address the safety, security and privacy concerns while advancing opportunities for innovation and utilization of UAV technology.}

{ In practice, depending on the size, weight, wing configuration, flying duration,  etc.,  UAVs  can be classified into different categories, such as large UAV versus small/mini UAVs, fixed-wing versus rotary-wing UAVs, etc. \cite{fotouhi2018survey}. Each type of UAVs generally possesses a set of unique characteristics and thus may be suitable for different application scenarios.  For example, fixed-wing UAVs have higher maximum flying speed, greater payloads, and longer flying endurance than rotary-wing UAVs, whereas the former requires a runway or launcher for takeoff/landing and it is also difficult for them to hover at a fixed position. In contrast, rotary-wing UAVs not only can take off/land vertically but also remain static at desired hovering locations, which renders them appealing for applications such as monitoring.  A detailed overview on different UAV classifications and applications was provided in \cite{hassanalian2017classifications}.}


Wireless communication is an essential technology to unlock the full potential of UAVs in numerous applications and has thus received unprecedented attention recently \cite{zeng2019accessing,mozaffari2018tutorial,fotouhi2018survey,zeng2020uavbook}.
As shown in Fig. 1, UAV communication technologies have evolved from the early direct link to the most recent 5G technologies in the course of the past twenty years, with significantly enhanced communication rate.  Although technologies such as direct link,  WiFi, and satellite communications are still useful in some remote scenarios where cellular services are unavailable, it is believed that exploiting the thriving 5G-and-beyond cellular networks to support UAV communications is the most promising and cost-effective approach, especially when the number of UAVs grows dramatically.
On the one hand, to guarantee safe and efficient flight operations of multiple UAVs, it is of paramount importance to provide secure and ultra-reliable communication links between the UAVs and their ground pilots or control stations for conveying command and control signals, especially in BVLOS scenarios.  Moreover, some practical UAV applications (e.g., real-time drone video filming and streaming to ground entities) require very high data rates  in the air-to-ground payload communication links. Fortunately, these requirements can in principle be largely met by cellular networks, thanks to their densely deployed communication infrastructure as well as advanced 5G-and-beyond technologies. On the other hand, because of  advances in communication equipment miniaturization as well as UAV manufacturing, mounting compact and lightweight base stations (BSs) or relays on UAVs becomes increasingly feasible. This leads to new types of flying aerial platforms that can be exploited to improve the quality of services for terrestrial users as well as to satisfy the need for on-demand deployment to address, e.g., temporary or unexpected events. This has led to two promising research paradigms for UAV communications, namely, UAV-assisted cellular communications and cellular-connected UAVs  \cite{zeng2019accessing}, where UAVs are integrated into cellular networks as aerial communication platforms and aerial users, respectively. As such, integrating UAVs into cellular networks is believed to be a win-win technology for both UAV-related industries and cellular network operators, which not only creates plenty of new business opportunities but also benefits the communication performance of  three-dimensional (3D) wireless networks.

\subsection{Communication, Sensing, and Intelligence in 3D UAV Networks}
According to  International Mobile Telecommunications-2020 (IMT-2020), 5G networks are expected to support a diverse variety of use cases in three main pillar categories \cite{shafi20175g}:
{\begin{itemize}
  \item {\it Enhanced mobile broadband (eMBB)} which aims to support data-intensive use cases such as  virtual and augmented reality (V/AR), requiring high data rates across a wide coverage area \cite{popovski20185g}.
  \item  {\it Ultra-reliable and low-latency communications (URLLC)} which aims to support mission-critical applications such as remote surgery, autonomous vehicles, and the Tactile Internet, requiring both ultra-high reliability and short delay \cite{ji2018ultra,8630650}.
  \item  {\it Massive machine-type communications (mMTC)} which aims to provide connectivity for massive numbers of power-limited devices with heterogenous traffic profiles such as in the industrial Internet-of-Things (IoT) \cite{bockelmann2016massive}.
\end{itemize}}
Although some of the above scenarios have been well studied for terrestrial wireless networks \cite{shafi20175g}, their techniques and results may not be directly applicable  to  future 3D wireless networks featuring  both UAVs (either as aerial users or as communication platforms) and ground users, due to their significantly different operating environments. More importantly, the new degrees of freedom they introduce for system design have been unexplored previously but can help further enhance the communication performance.

On the one hand, providing these services simultaneously for both UAVs and ground users imposes new technical challenges \cite{zeng2018cellular,zeng2019accessing,mei2019cellular,1080}. For example, since UAVs usually operate at much higher altitudes than conventional terrestrial users, exploiting the high beamforming gain via multiple antennas to provide eMBB service to UAVs requires ground BSs to have the capability of steering the beam not only in azimuth but also in the elevation plane. As such, the typical downtilted BS antennas need to be reconfigured, since currently they cater only to ground coverage as well as suppression of inter-cell interference in  long-term evolution (LTE) systems. In addition, the line-of-sight (LoS) dominated air-ground channels inherent to UAVs at high altitudes lead to smaller path loss due to less severe shadowing and multi-path fading. As a result, leveraging spectrum sharing to simultaneously support both UAVs and ground users in mMTC scenarios will generate significant ground-air interference to UAVs in the downlink, while in the uplink, UAVs will cause stronger interference to a large number of ground users even if they are distributed in the network \cite{mei2019cellular}.  Moreover, high UAV mobility generally results in more frequent handovers and time-varying wireless backhaul links between UAVs and ground BSs/users, which poses practical challenges in URLLC  scenarios. As such, it is imperative to develop new techniques and/or new designs, e.g., efficient handover management and resource allocation, to tackle the above issues to guarantee seamless service provisioning in the 3D space.

On the other hand, the emerging 3D wireless networks also introduce new design opportunities to better serve users in a cost-effective manner \cite{649,1074,JR:wu2018joint,zeng2016throughput}. For example,  by leveraging mobility to increase the operating altitude or bypass obstacles,  UAV-BSs/users are able to intelligently reposition themselves to avoid signal blockage to ground nodes caused by e.g., high-rise buildings. This is practically appealing for millimeter wave (mmWave) communications that are expected to provide eMBB services. Furthermore, for IoT/sensor networks, a hybrid network architecture composed of high altitude UAV-BSs/relays and their ground counterparts is able to achieve wider ground coverage. In addition, UAV-enabled mobile data collectors can move close to IoT devices to collect data, such that their transmit energy is minimized. This is particularly beneficial in mMTC scenarios where connectivity and energy consumption are critical factors \cite{Magazine:MTC}. Moreover,  non-trivial  tradeoffs may exist in leveraging these advantages  \cite{1074,wu2018common,yang2018energy,lyu2016cyclical}. For instance, although exploiting the high mobility of a UAV access point/data sink via trajectory design could improve the communication throughput, this comes at the cost of exceedingly long user access delays \cite{lyu2016cyclical,wu2018common}, which thus may not be appropriate for delay-sensitive URLLC applications.  Furthermore, to reap the promised benefits of UAV-assisted  cellular communications, there are still many important problems that need to be addressed, including more accurate air-ground channel modelling, traffic-adaptive UAV deployment and/or trajectory design, energy consumption modelling and energy-efficient design, and high-speed and reliable backhauling.

\begin{table*}[]
\centering
\renewcommand{\arraystretch}{1.6}
\caption{List of main industry progress, prototypes, and projects related to UAV communication networks.}\label{Table:industry:UAV}
\begin{tabular}{|c|c|p{12cm}|}
\hline
\textbf{Company}          & \textbf{Year}       & \multicolumn{1}{c|}{\textbf{Main activity and achievement}}                                                                                                                                                   \\ \hline
Qualcomm                  & 2016                & Feasibility proof of drone operation over commercial LTE networks at up to 400 feet (122 m) \cite{qualcom2016}.                                                                                                                     \\ \hline
Intel and AT\&T           & 2016                & Demonstrate the world's first LTE-connected drone at the 2016 Mobile World Congress \cite{intel2016}.                                                                                                                      \\ \hline
Ericsson and China Mobile & 2016                & Claim the world’s first 5G-enabled drone prototype field trial in WuXi of China \cite{Ericsson2016}.                                                                                                                                                                                                    \\ \hline
Nokia                     & 2016                &Design the flying-cell (F-Cell) which enables highly efficient ``drop and forget'' small cell deployments \cite{nokia2016}.                                                                                                                                                                                                        \\ \hline
Verizon                   & 2017                & Conduct flight tests using a ``flying cell site'' aboard a drone to supply an LTE network if severe weather knocks out more traditional cellular network infrastructure \cite{Verizon2017}.                                                                                               \\ \hline
Facebook                  & 2017                & Test ``Tether-Tenna'' to beam down the Internet from helicopters \cite{FB2017}.                                                                                                                                                                                                            \\ \hline
Huawei                    & 2019                & Announce 5G SkySite that contains a 5G remote radio unit (RRU) to provide signal coverage to 20-30 km$^2$ area  while flying at 100 metres \cite{Huawei2019}.                                                                                                                                                                                                          \\ \hline
\textbf{Research project} & \textbf{Start year} & \multicolumn{1}{c|}{\textbf{Main objective}}               \\ \hline
PercEvite                & 2017                &{Develop a lightweight and energy-efficient sensor, communication, and processing suite for small drones for autonomously detecting and avoiding ``ground-based'' obstacles and flying objects \cite{percevite2017,minucci2020avoiding}.}  \\ \hline
DroC$^2$om                 & 2017                &{Design and evaluate an integrated cellular-satellite system architecture for datalinks so that command and control information can be reliably transferred in support of functions and specific procedures for airspace access  \cite{dronecom2017,nguyen2018ensure}. } \\ \hline
SECOPS                 & 2017                &{Define an integrated security concept for drone operations that ensures that security risks are mitigated to an acceptable level \cite{SECOPS2017}.} \\ \hline
ABSOLUTE                  & 2017                & {Leverage  aerial BSs along with terrestrial and satellite communications to enhance the ground network capacity, especially for public safety in emergency situations \cite{absolute2017,912}.}  \\ \hline
DARE               & 2017                &{ Conduct advanced research on a new distributed autonomous and resilient emergency management system based on  wireless sensor, ad-hoc, and cellular networks\cite{DARE2017,deepak2019overview}.}     \\ \hline
5G!Drones        & 2019                & {Test use cases for vertical industry applications (IoT, industry 4.0, autonomous cars, etc.) on 5G test platforms \cite{5Gdrone2019,9303396}. }                                \\ \hline
AERPAW       & 2020                & {Build an aerial wireless experimentation platform spanning 5G technologies and beyond and enable cutting-edge research with the potential to create transformative wireless advances for aerial systems \cite{AERPAW2020,marojevic2020advanced}.}                              \\ \hline
\end{tabular}
\end{table*}

Besides the requirement for high-performance wireless communications, the ability to support effective and efficient \emph{sensing} is also essential for realizing the vision of integrating UAVs into 5G-and-beyond networks. On the one hand, sensing techniques are helpful in achieving safe UAV operation and intelligent air traffic management. On the other hand, UAVs can be leveraged as aerial sensing platforms to collect information in the sky. Recently, radio-based sensing technologies have received increasing attention due to their ability to achieve contactless and privacy-preserving object detection and environmental monitoring. Although the principles and objectives of radio-based sensing and radio-based communication systems are different, there have been significant research efforts to investigate their coexistence, cooperation, and joint design, thus leading to a new paradigm referred to as joint communication and sensing (JCAS) \cite{1234,1235,1241,1242}. This is because the same wireless infrastructure, RF hardware, and spectrum can be shared by both systems, which avoids the high costs of building dedicated wide-area sensing  infrastructure and also helps unleash the maximum potential of cellular networks. Despite these appealing advantages,  the research on JCAS in UAV communication networks is still in an early stage and there are many interesting and important problems that are open and require careful investigation.

Recently, artificial intelligence (AI) has been considered as another key enabling technology for 5G-and-beyond wireless networks integrated with UAVs \cite{Letaief_AI_Wireless2019,Gunduz_JSAC_2019}. This is mainly  due to its great potential to efficiently address challenging problems involving large amounts of data in system design and optimization, which positions AI-based approaches as powerful tools to facilitate highly dynamic UAV communication networks. For example, traditional off-line and model-driven trajectory design approaches usually require accurate and tractable communication models with perfect global knowledge of all  system parameters. This limits their applicability in practical scenarios with open operating environments and time- and spatial-varying traffic, especially when considering the real-time movement of UAVs and ground users in 3D space. Instead, by leveraging, e.g., deep reinforcement learning, UAVs can be endowed with the capability of predicting future network states in an online manner and thus adapt the communication resource allocation as well as the UAV trajectories based on the network  dynamics. Furthermore, UAVs with AI-embedded systems are also helpful in emerging applications such as edge computing, where multiple UAVs could work collaboratively either as aerial edge  servers or edge devices to achieve efficient data/computation  offloading.

\begin{table*}[]
\centering
\renewcommand{\arraystretch}{1.6}
{\caption{List of main 3GPP standardization progress on UAV communication networks \cite{muruganathan2018overview,950,3gppuav}.}\label{Table:industry}
{\begin{tabular}{| c | c |l|}
\hline
\textbf{3GPP Release No. }& \textbf{Study/Working Item}\\  
\hline
Release 15, TS 36.331 &Enhanced LTE Support for Aerial Vehicles\\  
\hline
Release 16, TS 22.125 & 	 Remote Identification of Unmanned Aerial Systems \\ 
\hline
Release 17, TS 22.125 & 5G Enhancement for UAVs\\
\hline
Release 17, TR 23.755 &  Study on Application Layer Support for Unmanned Aerial System (UAS)    \\
\hline
Release 17, TR 23.754& Study on Supporting Unmanned Aerial Systems Connectivity, Identification, and Tracking \\
\hline
\end{tabular}}}
\end{table*}

\subsection{ Industry Progress, Projects, and Standardization}

Recently, UAV communications have also drawn significant attention from industry, as shown in Table \ref{Table:industry:UAV}. Qualcomm's  field test was followed by a  trial report released in May 2017. The main focus of the report are the evaluation of the downlink signal-to-interference-plus-noise ratio (SINR) distribution and the study of the impact of power control and resource partitioning enhancements on uplink interference and throughput as the number of UAVs in the network grows.
Nokia's  flying-cell (F-Cell) is an experimental small cell that does not need any physical wires. In particular, a UAV obtains power from the surface-mounted solar panels and communicates with the carrier's core network over a high-speed wireless link, thus overcoming the challenges in backhaul cabling, deployment, and expenses.  While Qualcomm, Intel, AT\&T, Nokia, and Verizon used existing 4G/LTE cellular networks, Ericsson, China Mobile, and Huawei have conducted UAV experiments by leveraging more advanced 5G technologies. In particular, Huawei designed a 5G BS in 2019, called SkySite, which is a UAV mounted remote radio unit. Different from Nokia's F-Cell, Huawei's SkySite has a tethering  cable and it is used to provide a stable power supply as well as ultra-high speed and secured backhaul from the ground, thus enabling unlimited endurance with high communication performance.

There have also been several pilot projects launched in recent years to advance UAV research and  UAV field tests, as also shown  in Table \ref{Table:industry:UAV}.   In particular, in the ABSOLUTE  project, UAVs are leveraged as  aerial BSs for terrestrial and satellite communications to enhance the ground network capacity and extend the signal coverage, especially for public safety applications in emergency situations. Besides, the 5G!Drones project was  funded by the European Commission in 2019 to drive UAV vertical application trials by leveraging advanced 5G features, and involves 20 industrial and academic  partners from 8 European countries,  covering verticals, commercial network operators, networking industry, research centers, and universities. Specifically,  this project exploits network slicing as the key component to simultaneously run the three types of UAV services (eMBB, URLLC, and mMTC) on the same 5G infrastructure, demonstrating that each UAV application runs independently and does not affect the performance of other UAV applications, while providing different 5G services.

Besides, the third-generation partnership project (3GPP) has made significant efforts to ensure that cellular networks will meet the connectivity demands of UAVs \cite{muruganathan2018overview,950,3gppuav}, as shown in Table II.  In 2017, 3GPP approved a study item on enhanced  LTE support for UAVs, where the main objective was to identify  key challenges for using current LTE networks with  downtilted BS antennas to provide connectivity to UAVs. To further meet the requirements of  connectivity, identification, and tracking of UAVs, 3GPP recently considered the application of 5G networks in Release 17 \cite{3gppuav}. Several other standardization bodies and working groups have also devoted substantial efforts to develop different UAV specifications \cite{fotouhi2018survey}, including the International Telecommunication Union Telecommunication  (ITU-T) standardization sector, the European Telecommunications Standards Institute (ETSI), and the IEEE Drones Working Group (DWG). {Furthermore, to foster the research and innovation surrounding the study, design, and development of aerial communications,  IEEE Vehicular Technology Society (VTS) created an ad hoc committee on drones and
 IEEE Communication Society (ComSoc) established an emerging technology initiative  focusing on aerial users and networks  \cite{UAVeti,VTSdrone}.} One of the target applications of the  initiative is public safety, i.e., using aerial communications to deliver additional cellular coverage, for example during emergency situations, or to help first responders by providing advanced services. Moreover,  two standards working groups jointly sponsored by IEEE ComSoc and IEEE VTS focus on aerial communications and networking standards \cite{UAVVTS}.

 \subsection{Objectives and Contributions}


 {It is worth noting that  a number of papers in the literature provide  overviews or surveys on UAV communication research   \cite{649,618,949,913,967,952,950,khawaja2018survey,khuwaja2018survey,frew2008airborne1,bekmezci2013flying,gupta2016survey,hayat2016survey,
shakhatreh2018unmanned,motlagh2016low,shakeri2018design,mozaffari2018tutorial,fotouhi2018survey,zeng2019accessing,zeng2018cellular}. Unlike these works, this paper explicitly focuses on identifying the new major challenges posed by the  eMBB, URLLC, and mMTC use cases to UAV communication in 3D space, and  highlights the most promising solutions to tackle them.  Moreover, sensing and AI technologies are considered to further enhance the capabilities of future UAV cellular networks,  but were not covered in previous overview and survey papers.    To facilitate and inspire future research, various promising  directions for further investigation are also highlighted.}

Next, we provide an overview on the aforementioned five research areas respectively from Section II to Section VI. Specifically, in Section II, we highlight how emerging technologies, such as intelligent reflecting surface (IRS), can be leveraged to support eMBB applications in 3D space and also discuss the potential of exploiting massive multiple-input multiple-output (M-MIMO) and mmWave. In Section III, we provide an overview on short-packet communications (SPC) to guarantee URLLC in UAV networks.  In Section IV, we consider non-orthogonal multiple access (NOMA), energy harvesting, and energy efficient designs for the support of the massive connectivity.  In Section V,  we focus on radio-based sensing in wireless networks with UAVs and introduce a joint UAV communication and sensing approach. In Section VI,  we review machine learning methods particularly for UAV trajectory and communications design, discuss the computation offloading design for UAVs with mobile edge computing (MEC), and consider UAV-based distributed edge machine learning. In each section, directions for future research are also provided. Finally, the paper is concluded in Section VII.

\section{Enhanced Mobile Broadband}
eMBB is a natural evolution of 4G/LTE networks and will provide higher data rates and system capacity than current mobile broadband services, while guaranteeing moderate reliability, e.g., packet error rates (PERs) on the order of $10^{-3}$ \cite{shafi20175g}. In particular, 5G cellular networks are expected to support a peak data rate of 10 Gbits/s for eMBB applications, which in principle is sufficient for supporting high-rate UAV communication applications such as real-time video streaming and data relaying. On the other hand, UAV-enabled aerial platforms provide new degrees of freedom to further enhance the communication throughput or satisfy throughput requirements in a cost-effective manner. For terrestrial cellular networks,  a variety of viable solutions have been proposed to meet the stringent eMBB service requirements  \cite{wu2016overview}, namely, enhance spectral efficiency by deploying massive active transmit/receive antennas and/or passive reflecting elements such as  M-MIMO and IRS;  exploit unused new spectrum such as mmWave communications; reduce the transmitter-receiver distance and improve frequency reuse such as  ultra-dense networks (UDNs) and device-to-device (D2D) communications \cite{QQWU_2016_smcell,zhang2016fundamental,QQWU_2017_D2D}.  In this section, we focus on the exploitation of first two paradigms in 3D UAV cellular networks,  with a particular focus on the newly emerging  cost-effective IRS. {Generally speaking, M-MIMO based BSs can be deployed for a long-range coverage with massive devices, whereas mmWave and IRS are more suitable for short-range coverage, especially when there are strong deterministic channel components, such as the LoS path. Whereas both M-MIMO and mmWave require the use of active antennas arrays and generally have higher weight,  hardware cost and energy consumption than IRS that uses passive reflecting elements. Besides, due to the passive reflection mechanism of IRS, it requires a relatively larger aperture  than M-MIMO and mmWave based active arrays and also imposes new challenges on the related passive IRS channel estimation, which requires larger installation space in practice and may result in higher computational complexity.}

\subsection{{M-MIMO and mmWave for UAV Communications}}
M-MIMO, as a key enabling technology in the current 5G standard, is promising for supporting cellular-connected UAV communications \cite{huang2020massive,zeng2019accessing,chandhar2017massive,chandhar2019massive,garcia2019essential,geraci2018understanding}. Equipped with full-dimensional large arrays, ground BSs can perform fine-grained 3D beamforming to mitigate  interference among high-altitude UAVs and low-altitude terrestrial users, and thus, achieve much higher network throughput. However, the  beamforming gain of M-MIMO critically depends on the availability and accuracy of  channel state information (CSI) at the ground BSs, whereas cellular-connected UAVs introduce new challenges. First, UAVs can cause severe pilot contamination to a large number of ground BSs due to their strong LoS channels, which cannot be resolved by existing pilot decontamination techniques designed solely for  terrestrial users. Second, the UAVs' high mobility  in 3D space renders efficient beam tracking for them a more challenging task than for terrestrial users and can incur excessive pilot overhead \cite{huang20203d,lu2020pilot}. Third,  practical implementations may adopt the  hybrid beamforming based  M-MIMO architecture to support a large group of coordinated UAVs or a UAV swarm. This will  further complicate the pilot contamination and beam tracking problems. To better address the  inter-cell air-ground interference to achieve enhanced network throughput, a more ambitious approach referred to as cell-free M-MIMO, which combines ideas from distributed/network MIMO and coordinated multipoint transmission, was proposed recently, where massive antennas distributed over a large geographical area are connected with a central processing unit (CPU)  \cite{ngo2017cell}. In this case, both UAVs and ground users are surrounded by multiple  BS antennas, rather than the conventional case where each BS is surrounded by multiple  users \cite{ngo2017cell}. Due to the LoS-dominant air-ground channels, such a user-centric architecture provides more degrees of freedom to exploit the macro diversity offered by the many distributed BSs. Nevertheless, key issues that remain to be solved include efficient centralized and distributed power control, low-complexity fronthaul/backhaul provisioning, and network scalability with respect to UAV swarms \cite{d2020analysis,datta2020power,d2019cell,bjornson2020scalable}.


Another promising approach to support rate-demanding eMBB services in 3D space is to leverage the enormous chunks of new spectrum available in the mmWave bands \cite{zhang2019research,zhang2019survey}.  Several inherent limitations of mmWave communications, such as high signal attenuation and high vulnerability to blockage,  can be alleviated by exploiting the UAV mobility. For example, UAVs acting either as aerial platforms or users can not only fly towards ground nodes to reduce the propagation loss, but can also intelligently adjust their trajectories to bypass surrounding obstacles to increase the probability of LoS paths.  However, similar to M-MIMO, mmWave communication requires the  use  of a large number of antennas and the channel coherence time is shortened due to the high UAV mobility and shorter wavelength signals.
 As such, fast channel variations will result in practice, which renders effective dynamic beam training and tracking techniques imperative. Some existing works proposed the use of movement prediction filters such as Kalman filters to track the time-varying UAV-ground channels \cite{zhao2018channel,zhou2018beam,yuan2020learning}. Besides, a fast beam searching algorithm was developed in \cite{xiao2016enabling} to track the UAV-to-ground channels based on a predetermined hierarchical codebook. Other important topics for future research include  low-complexity  spectrum management \cite{feng2018spectrum}, and high-speed and reliable backhaul design.

\begin{figure*}[t]
\centering
\hspace{-0.9cm}\includegraphics[width=0.95\textwidth]{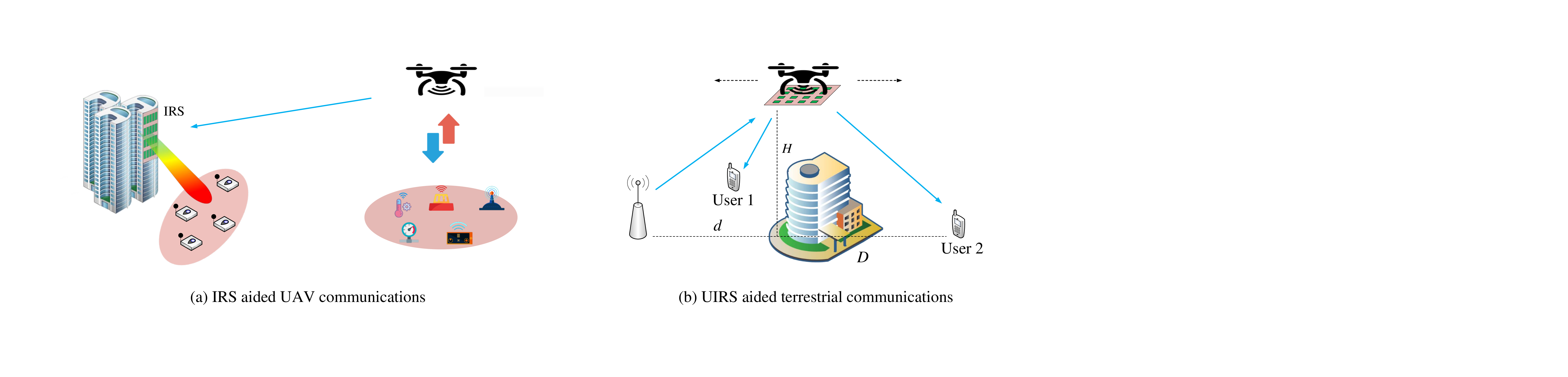}  
\caption{UAV communications with IRS. } \label{IRS:UAV}
\end{figure*}

\subsection{IRS for UAV Communications}
Despite the promising advantages of M-MIMO and mmWave communications,  their required high complexity and hardware cost as well as increased energy consumption are still crucial issues faced in practical implementation \cite{wu2016overview,zhang2016fundamental}.
{As an alternative, IRS has recently emerged as a new and  cost-effective solution to  improve the  received power and suppress air-ground interference in 3D space  \cite{JR:wu2019IRSmaga,gong2019towards,wu2020IRStutorial,basar2020reconfigurable}.} An IRS is composed of a large number of passive reflecting elements, each of which is able to reflect the impinging electromagnetic wave with a tunable reflection coefficient (including an amplitude and a phase shift) \cite{JR:wu2018IRS,di2020smart,JR:wu2019discreteIRS,huang2020holographic}. By smartly coordinating the reflections of all elements, an IRS is able to reconfigure the wireless channel with the desired signals added coherently and interference cancelled at designated receivers, thus significantly  enhancing the communication throughput without the need for deploying new active BSs or relays.
Furthermore, IRSs possess appealing advantages in practice such as light weight, and thus can be easily mounted on walls and even the surface of high-speed moving vehicles to support numerous applications \cite{zhang2020robust,lv2020secure,liu2020intelligent,yang2020intelligent,guan2020joint,}. As such, IRS has been considered as a disruptive technology for transforming our current ``dumb'' radio environment into an intelligent one, which potentially benefits a wide range of vertical industries such as transportation, manufacturing, and smart cities. Recently, IRS has also been recognized as a promising technology for the future sixth-generation (6G) ecosystem  \cite{rajatheva2020white,rajatheva2020scoring,wu2020IRStutorial} and studied in various system setups \cite{wu2019weighted,wu2021irs,chaccour2020risk,hua2020intelligent,guan2020intelligent,jamali2019intelligent,zheng2020intelligent,chen2019intelligent,wu2020joint,lu2020robust}. Generally speaking,  as shown in Fig.~\ref{IRS:UAV},  IRS can be either deployed on the ground to assist UAV communications or attached to UAVs to assist terrestrial communications \cite{lu2020aerial}, as elaborated below. {In addition, a brief comparison of existing works on IRS and UAV was summarized in Table \ref{tab:IRS:UAV}.}



\subsubsection{Terrestrial IRS Assisted UAV Communications}
 For  data dissemination and collection in IoT networks, a single UAV may need to fly sequentially approaching each IoT device/cluster for high data rate transmission. However, this not only compromises the user access delay but also increases the UAV propulsion energy consumption \cite{JR:wu2018joint,xu2018throughput,hua2019energy}. To overcome these drawbacks, one possible solution is to dispatch multiple UAVs, however this requires sophisticated UAV-UAV coordination and thus increases the operational cost as well as the signalling overhead. As an alternative, a properly deployed IRS can help resolve this issue. As shown in Fig. \ref{IRS:UAV} (a), the UAV only needs to directly cover a subset of the IoT devices and the significant path loss between the UAV and other devices can be effectively compensated by leveraging the deployment of IRS. This is particularly useful for uplink transmissions due to the limited energy supply of IoT devices, which helps reconcile the energy trade-off in ground-to-UAV communications \cite{yang2018energy}. In addition, as mentioned previously,  the ground BS antennas are  typically downtilted in current cellular networks, i.e., their main lobes point towards the ground for optimizing the coverage for ground users, and UAVs flying above the BSs are only supported through the side lobes. Indeed, recent 3GPP studies have confirmed that UAVs receive weak signals from existing terrestrial BSs and hence supporting aerial users will require further research and development \cite{muruganathan2018overview,950}. Fortunately, by leveraging the signals  reflected by IRSs via 3D passive beamforming  to serve the UAVs in the sky, the communication links between ground nodes and UAVs can be greatly improved, which helps eliminate the need for significantly modifying  the  configurations of existing ground BS antennas.


Motivated by the above advantages of IRS, there have been a number of works in this research direction  \cite{li2020reconfigurable,cai2020resource,wang2020joint,ge2020joint,ma2020enhancing,yang2020performance}. For instance, assuming that the UAV-ground link is blocked, the authors in \cite{li2020reconfigurable} maximized the average achievable data rate in IRS-assisted UAV-ground communication systems by jointly optimizing the UAV's trajectory and the IRS phase shifts. It was shown that rather than flying towards the ground user, the UAV should fly towards the IRS  to  significantly improve the data rates by exploiting the reflections of the large IRS aperture. Yet, this study only focused on the case of a single user being served by a single-antenna UAV and the proposed solution was not applicable to more general multi-user systems. To address this issue,  resource allocation, including user scheduling, power allocation, UAV trajectory, and beamforming was optimized  in \cite{cai2020resource} to minimize the total power consumption of an IRS-assisted multi-user multi-antenna UAV communication system, subject to a given required minimum achievable rate of each ground user. It was shown that deploying an IRS can shorten the UAV flying trajectory without compromising the QoS, as compared to the case without IRS. This work was then further extended to multi-IRS scenarios \cite{ge2020joint,wang2020joint}. In particular, the overall weighted data rate taking into account the geographical fairness of all the users was maximized in \cite{wang2020joint} to jointly optimize the UAV's trajectory and the phase shifts of the reflecting elements of the IRSs. Instead of using conventional convex optimization methods, the authors in \cite{wang2020joint} proposed a  low-complexity deep Q-network (DQN)-based solution by discretizing the trajectory, which is useful for practical systems with discrete phase-shift control. Then, they further proposed a deep deterministic policy gradient (DDPG)-based solution to facilitate continuous phase-shift control.

\begin{table*}[]
 \centering
\renewcommand{\arraystretch}{1.6}
\caption{Comparison of existing works on IRS and UAV. }\label{tab:IRS:UAV}
\begin{tabular}{|l|c|c|l|l|}
\hline
\textbf{IRS use case}                                  & \textbf{Reference} & \textbf{No. of IRS} & \textbf{Design Objective}   & \textbf{Approach}             \\ \hline
\multicolumn{1}{|c|}{\multirow{5}{*}{Terrestrial IRS}} & \cite{li2020reconfigurable}            & Single              & Rate maximization           & AO and SCA                    \\ \cline{2-5}
\multicolumn{1}{|c|}{}                                 & \cite{cai2020resource}            & Single              & Power minimization          & AO, SCA, and Lagrange duality \\ \cline{2-5}
\multicolumn{1}{|c|}{}                                 & \cite{wang2020joint}           & Multiple            & Weighted rate maximization  & Reinforcement Learning        \\ \cline{2-5}
\multicolumn{1}{|c|}{}                                 & \cite{ge2020joint}            & Multiple            & Received power maximization & AO and SCA                    \\ \cline{2-5}
\multicolumn{1}{|c|}{}                                 & \cite{hua2020uav}            & Multiple            & BER minimization            & Penalty based algorithm       \\ \hline
\multirow{4}{*}{UIRS}                                  & \cite{lu2020aerial}            & Single              & Minimum SNR maximization    & Two-step method               \\ \cline{2-5}
                                                       & \cite{zhang2019reflections}           & Single              & Rate maximization           & Reinforcement learning        \\ \cline{2-5}
                                                       & \cite{long2020reflections}          & Single              & Secrecy energy efficiency   & AO and SCA                    \\ \cline{2-5}
                                                       & \cite{shafique2020optimization}           & Single              & EE maximization             & Fractional programming        \\ \hline
\end{tabular}
\end{table*}


  In \cite{ma2020enhancing}, the potential of IRS to enhance cellular communications for UAVs was investigated,  where the received signal power gain was analyzed as a function of the UAV height and various IRS parameters including its size, altitude, and distance from the BS. In particular,  it was shown that  for a UAV hovering at $50$ meters (m) above the ground, a $21$ dB received power gain can be achieved with a properly deployed IRS comprising $100$ passive elements based on the 3GPP ground-to-air channel models. Besides, due to the downtilted antenna pattern of the BSs, the optimal IRS deployment altitude was shown to decrease as the IRS-BS distance increases so as to effectively reflect the signals coming from the BS. Also, in \cite{yang2020performance}, an IRS was deployed to assist a UAV relaying system where the average capacity, outage probability, and average bit-error rate (BER) were analyzed.  It was found that if the  IRS is deployed between the source and the UAV relay, the optimal position of the UAV for maximizing the outage performance moves closer to the destination, which is different from the case without the IRS where the UAV relay needs to be deployed in the middle between source and destination.

\subsubsection{UAV IRS Assisted Terrestrial Communications}  
In practice,  small lightweight UAVs usually can support only a limited payload, and may not be able to carry bulky RF transceivers in BSs/relays. Besides,  such active nodes generally incur high energy consumption, which further aggravates the energy consumption of UAVs and limits endurance. Furthermore, in terrestrial wireless networks, the deployment of IRS is highly restricted by the availability of  the existing infrastructure such as building facades, lamp posts, and advertising boards. As such, unlike the case in Fig. \ref{IRS:UAV} (a), it may not be possible to find proper deployment locations for installing an IRS to establish the desired (virtual) LoS links between the transceivers. Compared to terrestrial IRS, a UAV-based IRS (UIRS) is more likely to have strong LoS links with the ground nodes due to the relatively higher altitude and mobility of UAVs, thus reducing the probability of a blockage between them \cite{lu2020enabling}. This is particularly important for high frequency mmWave and THz communication systems that are very sensitive to blockage \cite{zhang2019reflections}.   For example, as shown in Fig. \ref{IRS:UAV} (b), since an IRS only reflects the signals to its front halfspace, deploying an IRS on the building facade only helps to enhance the communication data rate of user 1 and the lack of coverage behind the building remains still unsolved. Although it is possible to connect user 2 to the BS via multi-hop IRS reflections if multiple IRSs are deployed, this requires properly located buildings in the surrounding environment as well as additional multi-IRS coordination and signal processing complexity.  In this case, deploying a UAV equipped with an IRS is practically appealing \cite{lu2020aerial}.
More importantly,  UAVs with high mobility are able to adapt their positions dynamically according to the changes in the communication environment and thus can maintain persistent (virtual) LoS links between transceivers.

\begin{figure}[!t]
\centering
\includegraphics[width=0.49\textwidth]{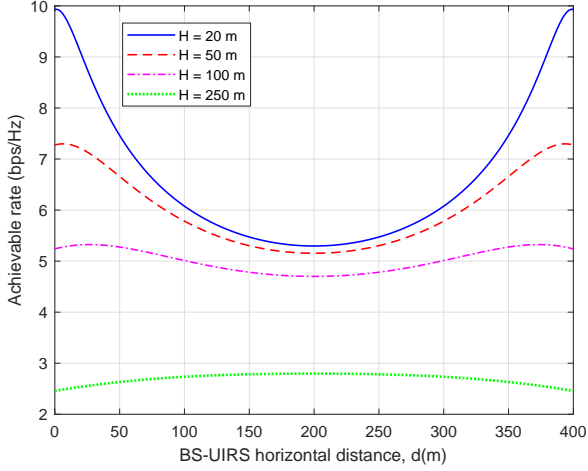}  
\caption{Achievable rate versus UIRS placement. } \label{IRS:UAV:placement}
\end{figure}

To provide some insights on UIRS deployment, in  Fig. \ref{IRS:UAV} (b), we focus on the UIRS-aided communication for user 2 and the BS-user distance is  $D$ m. The BS-UIRS horizontal distance is  $d$ m and the UIRS's altitude is $H$ m.  We optimize the horizontal location of the UIRS ($d$) to maximize the achievable rate/signal-to-noise ratio (SNR) of user 2. For simplicity, LoS channels are assumed for both the BS-UIRS and UIRS-user  links, while the BS-user direct link is assumed to be severely blocked and thus ignored. Then, the receive SNR at user 2 assuming optimal IRS passive beamforming  is given by \cite{JR:wu2019discreteIRS,lu2020aerial}
\begin{align}\label{rhoS}
{\mathrm{SNR}}=\frac{P\beta_0^2 N^2}{(d^2+H^2)((D-d)^2+H^2)\sigma^2},
\end{align}
where $P$ denotes the BS transmit power, $\beta_0$ denotes the channel power gain at a reference distance of $1$ m, $N$ denotes the number of IRS reflecting elements, and $\sigma^2$ denotes the receiver noise power. By setting the first-order derivative of the denominator with respect to $d$ to zero, the optimal UIRS placement for maximizing the receive SNR can be easily derived as \cite{JR:wu2019discreteIRS,lu2020aerial}
\begin{equation}\label{eq:scaling:law}
d^*=
\left\{
\begin{aligned}
&\frac{D\pm \sqrt{{D^2}-{4H^2}}}{2},  && \text{for}~~ D\geq 2H,  \\
&\frac{D}{2},  && \text{otherwise}.
\end{aligned}
\right.
\end{equation}
 It is interesting to note that depending on the values of $H$ and $D$, the optimal deployment location changes from the midpoint to  locations with equal distances to the transmitter (BS) and the receiver (user 2), as also verified  in Fig. \ref{IRS:UAV:placement} for $P=20$ dBm, $\beta_0=-30$ dB, $N=250$, $\sigma^2=-100$ dBm, and $D=400$ m.  Considering that $D$ in practice  is expected to be considerably  larger than $H$ to obtain effective reflection from the UIRS,  the UIRS should be deployed close to the transceivers, which is quite different from the deployment strategy for active relays. {Nevertheless, practical  UIRS deployment also needs to consider other factors such as LoS versus non-LoS links, channel rank and condition number, and the use of a single large UIRS or multiple smaller cooperative UIRSs \cite{JR:wu2018IRS,wu2020IRStutorial,zhang2020intelligent}. Furthermore, the physical UIRS elements orientation also has a great effect on the air-ground channel  characteristics and hence the system performance, which calls for further investigation in future.}

However, there has been only limited work in this line of research so far \cite{zhang2019reflections,shafique2020optimization,long2020reflections,jiao2020joint,lu2020enabling}. In \cite{lu2020enabling}, the minimum SNR within a given rectangular service area (e.g., a hot spot in a cellular network) was maximized by jointly optimizing aerial IRS placement, the IRS phase shifts, and the transmit beamforming at the BS. To gain useful insights, the authors first derived the optimal UIRS placement and phase shifts for the special case of single-location SNR maximization, where  the optimal horizontal location was shown to depend only on the ratio between the UIRS height and the source-destination distance. As for the general case of area coverage enhancement, an efficient two-step method was proposed by decoupling the phase-shift optimization from the UIRS placement design, where a 3D beam broadening and flattening technique was proposed to obtain 3D beam patterns matching the service area size on the ground \cite{lu2020aerial}. For mobile users in practical systems, the dynamic self-positioning of UIRS is required, which motivated the work in \cite{zhang2019reflections}. Specifically, to maximize the downlink mmWave communication capacity of a mobile outdoor user,  a reinforcement learning (RL) approach based on Q-learning and neural networks was proposed  in \cite{zhang2019reflections} to enable efficient deployment and phase-shift optimization for a self-sustained UIRS. The main advantage of this approach lies in that it requires no prior knowledge of the dynamic environment, but learns the characteristics of the environment during the service process of the UIRS, based on data measurements and feedback received during each communication stage. The results showed that significant gains can be achieved by using such a UIRS as compared to a static IRS in terms of the average data rate as well as the LoS probability. Yet, to simplify  the system design, it was assumed in \cite{zhang2019reflections} that the BS only transmits, and thus, the UIRS only reflects signals during the hovering state. 

\subsection{Future Research}

Despite these existing works, the study of IRS-assisted eMBB air-ground communications   is still in a very early stage and many interesting and important problems need to be addressed.

\subsubsection{Joint UAV Trajectory and IRS Reflection Pattern Design}
First, the IRS reflections could be very sensitive to UAV jittering and user location uncertainty \cite{xu2020multiuser}, which may cause significant performance degradation for UAVs and/or ground users due to the reliance on LoS channels. Particularly, the current joint UAV trajectory and IRS phase-shift designs utilized  trajectory discretization \cite{li2020reconfigurable,cai2020resource,ge2020joint,wang2020joint} where the UAV is considered to be at a fixed sample location during each interval. However, in reality,  the UAV is continually moving, which can easily render the IRS passive beamforming ineffective due to the signal misalignment arising from UAV movement. As such, the development of robust algorithms for tuning the phase shifts at the IRS  is still a practically challenging task. Joint reflection amplitude and phase-shift control at the IRS could be a possible approach to alleviate air-ground interference, which, however, increases the hardware cost and design complexity. Hence, how to strike an optimal performance-complexity/cost tradeoff when considering practical hardware imperfections such as discrete and/or coupled reflection amplitudes and phase shifts \cite{zhao2020intelligent,abeywickrama2019intelligent,wu2018IRS_discrete,zhao2020exploiting} remains an open problem. In addition, practical low-complexity channel estimation/tracking methods are required for acquiring the UIRS-user channels along their 3D trajectories, which is more challenging than the case of terrestrial IRSs that are at fixed locations \cite{wu2020IRStutorial}.

\subsubsection{Deployment of IRS/UIRS and UAV-IRS Symbiotic Systems}
For a given network with both ground users and UAVs, how to jointly optimize the resource allocation and the deployment of  terrestrial IRSs and UIRSs including their locations and densities is another important and interesting topic to pursue. Finally, most existing studies focus on the utilization of IRSs to enhance the primary end-to-end communication by exploiting passive beamforming \cite{li2020reconfigurable,cai2020resource,ge2020joint,wang2020joint,ma2020enhancing,yang2020performance,zhang2019reflections,shafique2020optimization,long2020reflections,jiao2020joint,lu2020enabling}. However, in practice, UIRSs also need to deliver their own information, including control signals to acknowledge their current status,  environmental parameters (such as temperature, humidity and pressure) obtained by UAV sensors, etc., to the BS.  Rather than equipping each UIRS with a dedicated transmitter, which is not cost-effective and also requires extra power consumption, a more appealing approach is to modulate the IRS information onto the reflected signals to achieve passive information transfer, which leads to the new paradigm of symbiotic communication \cite{hua2020uav,zhang2020large,yan2019passive} that aims to simultaneously transmit  IRS information and enhance the  communication quality of the primary link.

\begin{table*}[t]\caption{Applications and Requirements for URLLC in 5G-and-beyond Systems \cite{JR:URRLC_requirements_1,JR:URRLC_requirements_2,3gpp_release_16}. }\label{tab:URRLC_requirements} \centering
\renewcommand{\arraystretch}{1.6}
\begin{tabular}{|l|l|l|l|l|}\hline
Applications  & Latency (ms)&  Reliability (\%)& Data Size (bytes) & Communication Range (m)\\
                                                        \hline
Smart Grid & $3\sim20 $ & 99.999 & $80\sim 1000 $& $10\sim 1000 $ \\
    \hline
Professional Audio  &$2 $ & 99.9999 & $3\sim 1000 $& $100 $ \\
 \hline
Automated Vehicles  & $1$ & 99 & $144 $& $400 $ \\
    \hline
    E-Health  &  $30$ & 99.999 & $28\sim 1400 $& $300\sim 500 $ \\
        \hline

Argument Reality & $0.4\sim 2$ & 99.999 & $12,000\sim 16,000 $& $100\sim 400 $ \\
       \hline
   Intelligent Transportation Systems  & $10$ (end-to-end) & 99.9999  & $50\sim 200 $& $300\sim 1000 $ \\
       \hline
          Vehicle-to-vehicle (V2V) & $5$ & 99.999 & $1600$& $300 $ \\
       \hline
        Tactile Internet & $1$ & 99.99999 & $250 $& $100$k \\
       \hline
      Internet-of-Drones (Video) & $1\sim10$ & 99.999 & $1-20,000 $& Ground-to-air $\sim 400$  \\
       \hline
\end{tabular}
\end{table*}

%
\section{Ultra-reliable and Low-latency Communications}

To fully embrace the era of the  Internet-of-Everything (IoE), a well-developed wireless infrastructure focusing on low latency and high reliability is necessary. As a result, starting from the 5G wireless systems, the concept of URLLC  has been introduced as a key performance indicator  \cite{3gpp_release_16,book:Kwan_5G}. Unlike classical high data rate oriented multimedia streaming services,  in which high rate data  flows  from a source to a sink, typical URLLC services focus on conveying sensing information, control commands, and feedback information in short packets which need to be delivered reliably in an extremely short period. Table \ref{tab:URRLC_requirements} shows some emerging URLLC applications and requirements. It can be seen  that these applications impose heterogeneous requirements on both the reliability and latency.  In particular, for the Internet-of-Drones (IoD), the need for rapidly conveying control signals in highly dynamic UAV communication scenarios imposes strict requirements on both  latency and reliability. Unfortunately, despite the development of advanced wireless communication techniques, e.g.,
massive MIMO \cite{JR:JSAC_B5G_antennas,JR:EE_massive_MIMO_Kwan,JR:large_number_antennas,JR:EE_SWIPT_Massive_MIMO}, mmWave transmission \cite{JR:Lou_mmwave,JR:mmwave_Zhiqiang1,JR:mmwave_Zhiqiang2}, and full-duplex communication \cite{JR:Nguyen_MIMO_FD_EE,JR:Yan_MOOP,duo2020energy}, it is believed that a single wireless link between a user and a terrestrial BS cannot satisfy URLLC requirements. As a result, different kinds of diversities are necessary to offer the flexibility to trade more bandwidth for achieving shorter delays and higher reliability \cite{CN:Tony_URLLC}.

Recently, UAV-based communications have been widely recognized as one of the disruptive and enabling technologies to address this issue.  Indeed, aerial nodes usually enjoy unobstructed  LoS air-to-ground channels \cite{JR:UAV_channel_measurement}. Thus, conventional terrestrial communication systems with blocked communication links may deploy UAVs as aerial relays to substantially improve system data rate and/or coverage to realize URLLC by exploiting spatial diversity. Besides, in the case of natural disasters and disease outbreaks, it may not be possible to guarantee URLLC via  standalone conventional terrestrial communication networks. In these cases, UAVs can be employed as ad-hoc aerial BSs to offer temporary URLLC links. In fact, by exploiting the high maneuverability of UAVs, fast, highly flexible, and cost-effective deployment of communication infrastructure can be ubiquitously established, especially in temporary hot spots, disaster areas, complex terrains, and rural areas.  On the other hand, UAVs can also serve as a physical carrier for user equipment connecting to existing networks requiring URLLC services, e.g., to report sensing information.
In both cases, physical communication of UAVs can be divided into payload delivery (PD) and control and non-payload communication (CNPC) \cite{zeng2019accessing}. Specifically, for CNPC links, short-packet control information is required to be conveyed from a ground BS to UAVs or exchanged between UAVs with high reliability and low latency \cite{report:UAV_control_NASA}. As for PD links, UAVs can be adopted to provide certain types of data services requiring not only extremely high data rates but also ultra-low latency and high reliability, e.g. conveying real-time video and haptic information in the Tactile Internet. However, existing wireless communication protocols and techniques were designed mainly for terrestrial communications and do not fully capture the hybrid characteristics of air-to-ground and air-to-air communications. More importantly,  applying conventional techniques/protocols to UAV-based communication systems may introduce an exceedingly long delay and raises serious safety issues for controlling aerial nodes, which remains a major obstacle for realizing efficient UAV communications. Next, we show how to leverage short-packet communication to meet the requirement of URLLC applications.

\subsection{Short-Packet Communication}
To guarantee URLLC in UAV-based communication systems, various approaches have been proposed in the literature and 5G standardization. For instance, by exploiting the excessive spatial degrees of freedom offered by massive MIMO, in theory, the reliability of UAV-based communications in the ground-to-air links can be theoretically  improved to a certain extent \cite{JR:UAV_massive_MIMO}. However, the deployment of a large number of antennas at UAVs is challenging due to the aforementioned energy and size limitations of practical UAVs.  The actual communication latency, $ T_{\mathrm{L}}$, is comprised of several components which can be summarized as follows:
\begin{eqnarray} \label{eqn:end-to-end_delay}
T_{\mathrm{L}}= T_{\mathrm{ttt}}+ T_{\mathrm{prop}}+ T_{\mathrm{proc}}+ T_{\mathrm{retex}}+ T_{\mathrm{sig}},
\end{eqnarray} where $ T_{\mathrm{ttt}}$ denotes the time-to-transmit latency, which is the time needed to transmit a packet; $ T_{\mathrm{prop}}$ is the signal propagation delay from the transmitter to the receiver; $ T_{\mathrm{proc}}$ is the time consumed for encoding, decoding, synchronization, and channel estimation; $ T_{\mathrm{retex}}$ is the time required for retransmission (if any); and $T_{\mathrm{sig}}$ is the time incurred by signaling exchanges such as connection request, scheduling grant, channel training,  feedback, and queuing delay. In response to the harsh maximum delay allowance specified in Table \ref{tab:URRLC_requirements}, different methods for reducing the overall communication latency have been proposed in the literature \cite{Mag:URLLC:Frame_structure} by minimizing the different components in \eqref{eqn:end-to-end_delay}. For instance, if the communication link is shadowed, one can deploy a UAV as a wireless relay to establish an additional reliable  end-to-end communication link which can help to reduce the time spent on retransmission, i.e., $T_{\mathrm{retex}}$. Also, non-coherent transmission and detection were advocated in \cite{JR:Petar_URLLC_TCOM}. In particular, for these schemes, acquiring the knowledge of instantaneous channel conditions is not necessary, which reduces both $ T_{\mathrm{proc}}$ and $T_{\mathrm{sig}}$ needed for channel training and estimation at the expense of some performance loss in signal detection.  Besides, a new short slot structure exploiting the concept of mini-time slots was proposed for 5G \cite{JR:5G_mini_time_slot} to realize URLLC in practice via shortening $T_{\mathrm{ttt}}$ in \eqref{eqn:end-to-end_delay}.

On the other hand, to further improve system performance, channel capacity-achieving error-correcting codes have been proposed and adopted in 5G communication systems, e.g., polar codes and low-density parity-check (LDPC) codes. However, due to the requirement of short packets with low latency, the blocklength of the coded messages is short which in turn jeopardizes the communication reliability. More importantly, the Shannon capacity, which assumes infinite blocklength codes, is no longer achievable and cannot be used to accurately characterize the system performance.  Although one can perform proper resource allocation to reduce the performance degradation by taking into account the impact of short blocklength codes, most existing resource allocation algorithms for UAV communications, e.g., \cite{JR:Yan_UAV_TCOM,JR:Ruidei_UAV_TCOM,JR:Cai_UAV_TCOM,CN:Xiaoming_UAV}, were designed based on Shannon capacity as the performance metric, thereby assuming implicitly infinite block length transmission.  Besides, these algorithms do not offer sufficient flexibility to strike a balance between delay (codelength) and reliability.

Recently, driven by the need for URLLC, the study of SPC has received significant attention in the literature \cite{JR:Xixi_URLLC_UAV,JR:Yansha_URLLC_UAV,JR:Wei_Xu_URLLC_UAV,CN:Robert_URLLC_UAV}, {as summarized in Table \ref{tab:URRLC_techniques}.}  These works have captured the relationship between the achievable rate, decoding error probability, and packet length, which facilitates the design of effective resource allocation. In the following, we summarize the key techniques for resource allocation design taking into account the impact of short packets in UAV communications.

\subsubsection{Performance Metric for SPC}
Unlike conventional systems, where Shannon's capacity theorem can be used to characterize performance and the probability of erroneous decisions becomes negligible as the coded packet length is sufficiently long, short packets are adopted in URLLC systems to achieve low latency, which makes decoding errors unavoidable. As an alternative, for evaluating the performance of SPC, the notion of \emph{normal approximation} for finite blocklength codes was developed in \cite{JR:normal_approximation_original}. For ease of illustration, we focus on a simple UAV-to-ground BS communication channel. Suppose that the BS transmits a message to the  UAV over a quasi-static flat fading channel with a given packet length, $N_{\mathrm{B}}=WT$, where $W$ is the bandwidth and $T$ denotes the transmission duration. The achievable
rate\footnote{Based on the results in \cite{Mag:normal_approximation_gap}, the approximation error in \eqref{eqn:capaciity_approximation} is $0.1-2$ dB for decoding error probabilities from $10^{-3}$ to $10^{-7}$. } (bit/s/Hz)  for the given finite blocklength can be accurately approximated as follows \cite{JR:normal_approximation_original}:
\begin{eqnarray}\label{eqn:capaciity_approximation}
	R\approx C-\sqrt{\frac{V}{N_{\mathrm{B}}}}\frac{Q^{-1}(\epsilon)}{\ln 2},
\end{eqnarray}
where $C=\log_2(1+\gamma)$ denotes the Shannon channel capacity that can be achieved for infinite packet lengths,  $\gamma$ is the received SNR which is a function of the UAV's trajectory, and $\epsilon$ is the required decoding packet error probability. Variable $V$ is the channel dispersion and $Q^{-1}(\cdot)$ is the inverse of the Gaussian Q-function, where $
Q(x)=\frac{1}{\sqrt{2\pi}}\int_{x}^{\infty}\text{exp}{\left(-\frac{t^{2}}{2}\right)}\text{d}t$. For the complex-valued additive white Gaussian noise (AWGN) channel, the channel dispersion is given by
\begin{eqnarray}\label{dispersion}
	V=1-{(1+\gamma)^{-2}}.
\end{eqnarray}
As can be observed from  \eqref{eqn:capaciity_approximation}, the subtrahend accounts for the penalty caused by the finite blocklength. In particular, for a sufficiently long blocklength, the normal approximation approaches the conventional Shannon capacity. As a result,  the normal approximation in  \eqref{eqn:capaciity_approximation} serves as a useful tool for optimizing the allocation of the system resources to strike a balance between the blocklength (delay), achievable rate, and reliability.  In the following, we focus on an illustrative example to highlight how URLLC can be facilitated in UAV communication systems via resource allocation.

\subsubsection{Example for Resource Allocation for URLLC}
\begin{figure}[t]
\centering
\includegraphics[width=3.5  in]{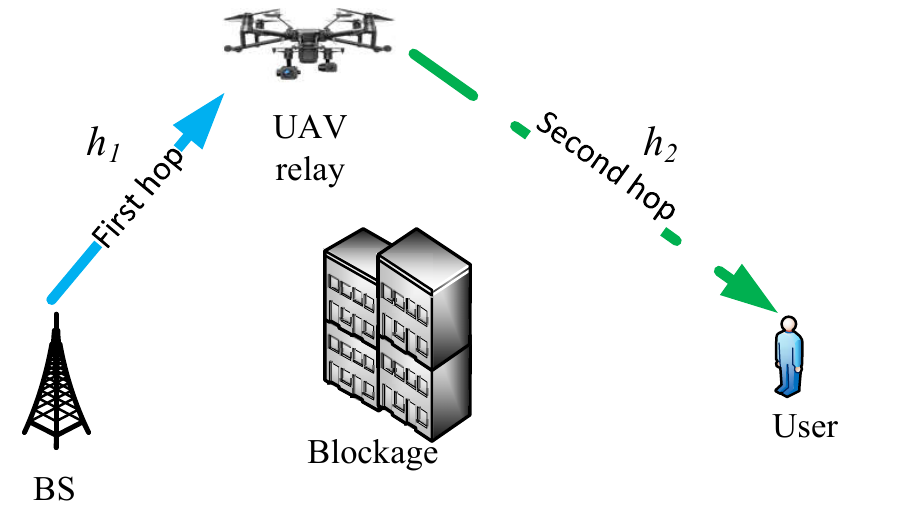}%
\caption{An example of adopting a UAV as a wireless relay taking into account the impact of finite blocklength. } \label{fig_block_length_UAV}
\end{figure}

 Suppose there is a ground BS communicating with a ground user. Due to heavy shadowing in the direct channel, as shown in Fig. \ref{fig_block_length_UAV},  a UAV adopting the decode-and-forward protocol is dispatched to serve as a wireless relay to facilitate  reliable end-to-end communication resulting in a two-hop relaying system \cite{CN:UAV_relaying_Xiaofang,JR:UAV_relaying}. The 3D coordinates of the BS, the user, and the UAV are $(0,0,0)$, $(a,b,0)$, and $(x,y, H)$,  respectively,  with $H$ being the UAV's fixed altitude. For ease of presentation, we denote the channel power gain from the BS to the UAV and that from the UAV to the user by $h_1$ and $h_2$, respectively. Also, the transmit powers of the BS and the UAV are given by $P^{\mathrm{BS}}$ and $P^{\mathrm{UAV}}$, respectively. It is well known that for a sufficiently high altitude, the free-space channel model can accurately capture the channel characteristic. Thus, $h_1$ and $h_2$ are modelled as
\begin{eqnarray}
\hspace*{-1mm}h_1=\frac{\beta_0}{H^2+x^2+y^2} \,, \, h_2=\frac{\beta_0}{H^2+(a-x)^2+(b-y)^2},
\end{eqnarray}
respectively, where $\beta_0$ is the channel power gain at a given reference distance.  As a result, the received SNR at the UAV and the user can be expressed as
\begin{eqnarray}\label{eqn:SNR_URLLC}
\gamma_1=\frac{P_{\mathrm{BS}}h_1}{\sigma_{\mathrm{UAV}}^2}\,,  \,
\gamma_2=\frac{P_{\mathrm{UAV}}h_2}{\sigma_{\mathrm{User}}^2},
\end{eqnarray}
respectively, where $\sigma_{\mathrm{UAV}}^2$ and $\sigma_{\mathrm{User}}^2$ denote the power of the AWGN at the UAV and the user, respectively.

As can be observed from \eqref{eqn:SNR_URLLC},  the received SNRs of the two hops depend on the position of the UAV. Then, by applying the normal approximation in \eqref{eqn:capaciity_approximation}, the achievable rates of the first hop and the second hop are given by
\begin{eqnarray}
R_1(m_1,x,y)=\log_2(1+\gamma_1)-\sqrt{\frac{V_1}{m_1}}\frac{Q^{-1}(\epsilon)}{\ln 2},\\
R_2(m_2,x,y)=\log_2(1+\gamma_2)-\sqrt{\frac{V_2}{m_2}}\frac{Q^{-1}(\epsilon)}{\ln 2},
\end{eqnarray}
respectively, where $m_1$ and $m_2$ are the blocklengths adopted at the BS and the UAV, respectively. Furthermore, the channel dispersions of the first hop and the second hop are given by $	V_1=\bigg(1-{(1+\gamma_1)^{-2}}\bigg)$ and 	$V_2=\bigg(1-{(1+\gamma_2)^{-2}}\bigg)$, respectively.

In practice, it is common to maximize the total achievable rate of the user  by jointly optimizing the position of the UAV relay and the blocklengths in the two hops, taking into account the end-to-end communication delay requirement. Thus, the proposed resource allocation policy is determined by solving the following optimization problem:
\begin{eqnarray}\label{eq:optimization _finite_block_length}
 ~~\underset{{m_1, m_2, x, y}}{\maxo} ~~~&& \frac{1}{2}\min\Bigg\{R_1(m_1,x,y), R_2(m_2,x,y) \Bigg\} \notag \\
\mathrm{s.t.}~~~~&& \mbox{C1: } (m_1 + m_2)T_{\mathrm{Block}}\leq D_{\max},\label{power:constraints}\notag   \\
&& \mbox{C2: }m_i\in \mathbb{Z}^+, i\in{1,2},  \label{blocklength:constraints}\notag \\
&& \mbox{C3: }x^2+y^2\le r^2_{\max},
\end{eqnarray}
where $T_{\mathrm{Block}}$ in  \eqref{eq:optimization _finite_block_length} denotes the time needed for conveying one unit of blocklength and constraint $\mbox {C1}$ guarantees that the maximum transmission delay does not exceed $D_{\max}$. Constraint $\mbox {C2}$ enforces that the blocklengths are non-negative integers.
Finally, constraint $\mbox {C3}$ ensures that  the position of the UAV relay is within the serving cell with a radius of $r_{\max}$.

Problem  \eqref{eq:optimization _finite_block_length} is a non-convex optimization problem. The non-convexity is caused by the combinatorial constraint in \mbox{C2} and the non-convex normal approximation appearing in the objective function.
However,  by analyzing the structure of problem \eqref{eq:optimization _finite_block_length},  it can be solved optimally via a combination of the penalty method and monotonic optimization \cite{ghanem2019resource} at the expense of a high computational complexity. As an alternative, one can apply approximations to simplify the problem at hand. For instance, a high SNR approximation is commonly adopted. In particular, in the high SNR regime, i.e., $\gamma \rightarrow \infty$, the channel dispersion can be approximated by
\begin{eqnarray}
V=1-\frac{1}{(1+\gamma)^2}\approx 1
\end{eqnarray}
such that it becomes a constant. Then, successive convex approximation and the penalty method can be employed to obtain a suboptimal solution to \eqref{eq:optimization _finite_block_length} with polynomial time computational complexity.
 \begin{figure}[t]
 \centering
\includegraphics[width=3.5 in]{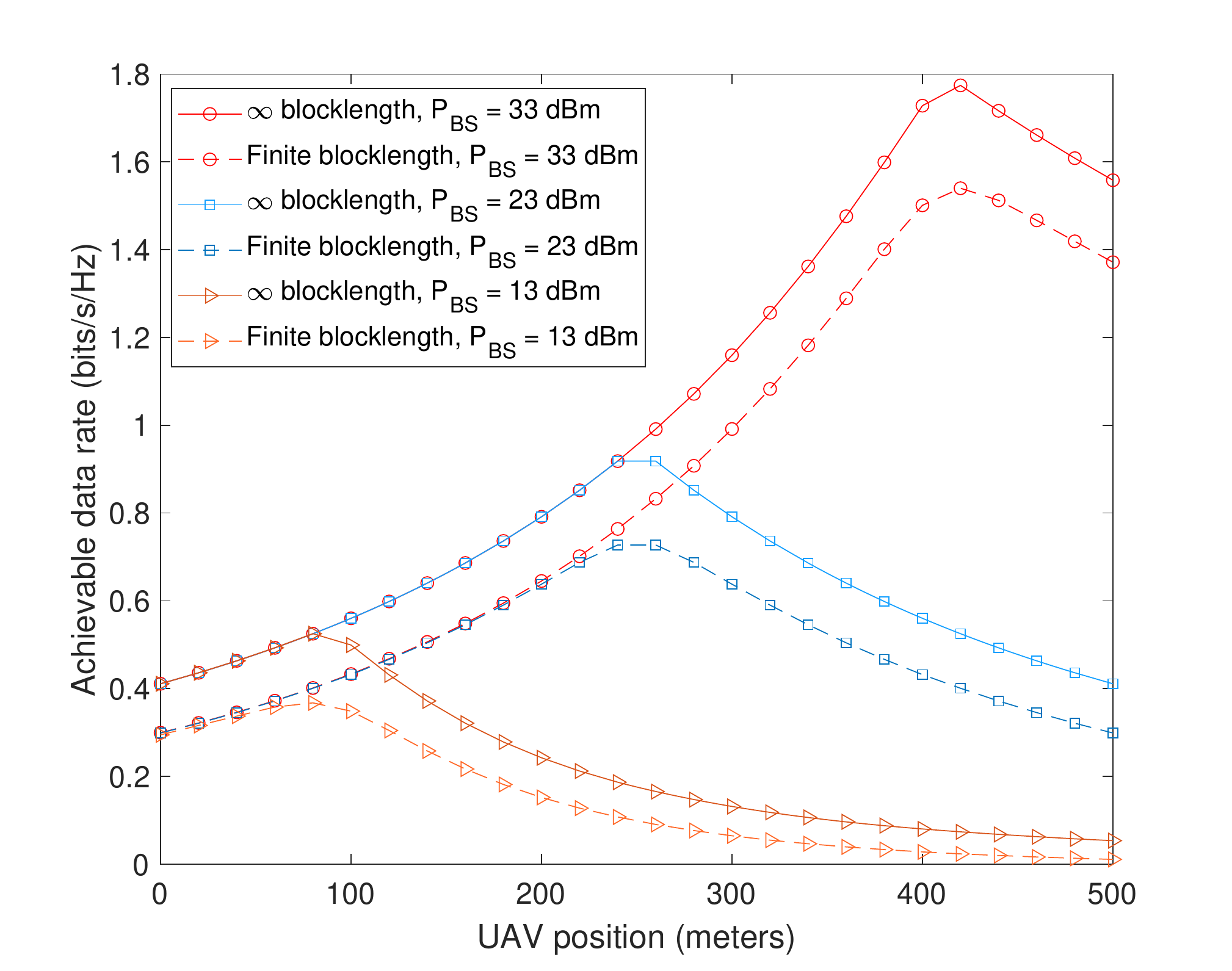}
 \caption{A comparison of the achievable data rate of a UAV-assisted relaying network for different BS transmit powers and different blocklengths. The transmit power of the UAV hovering at a fixed altitude $H=100$ m is $23$ dBm. A ground user is located  $500$ m away from the BS. } \label{fig:URLLC_simulation}
\end{figure}

In Fig. \ref{fig:URLLC_simulation}, we show the achievable data rate of the considered UAV-assisted relaying system versus the position of the UAV for different BS transmit powers. The maximum delay is $1$ ms while each mini-time slot is $T_{\mathrm{Block}}=0.01$ ms with $200$ kHz of bandwidth. In other words, there are a maximum of $100$ blocks for transmission. For simplicity, the channel-to-noise ratio (CNR) at the reference distance of $1$ meter is set as $\frac{\beta_0}{\sigma_{\mathrm{User}}^2}=\frac{\beta_0}{\sigma_{\mathrm{UAV}}^2}=60$ dB and the required decoding packet
error probability is $\epsilon=0.1\%$. The achievable rate for the finite blocklength case is obtained by optimizing $m_1$ and $m_2$ in \eqref{eq:optimization _finite_block_length}. For comparison, we also consider a performance benchmark for which the delay constraint C1 in \eqref{eq:optimization _finite_block_length} is dropped and an infinite blocklength is assumed for transmission. As can be observed from Fig. \ref{fig:URLLC_simulation}, there is a performance gap between the cases of finite blocklength and infinite blocklength, as the former has an insufficient number of coded information blocks for sufficiently averaging out  the impact of Gaussian noise to approach the system capacity.  This gap is generally inevitable in URLLC as the end-to-end delay constraint is stringent.  Furthermore, the optimal UAV position  maximizing the end-to-end  achievable rate depends heavily on the power budgets of both the UAV and the ground BS. In fact, the optimal UAV position attempts to balance the achievable rate of the two hops in the system. In particular, when the power budget at the BS is smaller than that of the UAV, e.g. $P_{\mathrm{BS}}=13$ dBm, $P_{\mathrm{UAV}}=23$ dBm, the optimal UAV position is closer to the BS than to the user. Although the power budget of the BS is limited, the optimal strategy reduces the path loss between the BS and the UAV for establishing an efficient first communication hop. Besides, the higher UAV transmit power  can be exploited to compensate for the longer communication distance between the UAV and the ground user to balance the achievable rate of the two hops. In contrast, when the power budget of the BS is larger than that of the UAV, e.g., $P_{\mathrm{BS}}=33$ dBm and $P_{\mathrm{UAV}}=23$ dBm, the optimal UAV position moves towards the user, allowing the system to again fully exploit the power budgets of both the BS and the UAV to maximize the system performance.

\begin{table*}[t]
\caption{Comparison of existing works on URLLC in UAV communications. }\label{tab:URRLC_techniques} \centering
\renewcommand{\arraystretch}{1.0}
\begin{tabular}{|l|p{3cm}|p{4.5cm}|p{3cm}|p{4cm}|}\hline
Reference  & Methodology&  Objective & Advantage & Disadvantage\\

                                                        \hline
\cite{JR:Petar_URLLC_TCOM} & Non-coherence transmission and detection & Reducing $ T_{\mathrm{proc}}$ and $T_{\mathrm{sig}}$   & Easy to implement& Sensitive to decoding error \\
    \hline
\cite{JR:5G_mini_time_slot}  & Introducing mini-time slot & Shortening $T_{\mathrm{ttt}}$ & Achieving low latency& Protocol update is needed \\
    \hline

\cite{8594709} & Short error-correcting codes & Reducing $ T_{\mathrm{proc}}$ and $T_{\mathrm{sig}}$   & Striking a balance between latency and error rate & Difficult to design efficient code structure and  decoder \\
    \hline
  \cite{JR:Xixi_URLLC_UAV}  &  Joint UAV altitude and transmission duration optimization & Maximizing available UAV communication range subject to delay constraints & Enjoying LoS and adaptive to actual environment&  Challenging optimization problem\\
        \hline

\cite{JR:Yansha_URLLC_UAV} & Joint blocklength and UAV variable optimization & Minimizing end-to-end communication error probability subject to blocklength constraints & Enjoying LoS and adaptive to actual environment & Challenging optimization problem\\
       \hline

\cite{CN:Robert_URLLC_UAV} & Resource allocation for short packet communication & Maximizing the achievable rate for a given finite blocklength  &  Adaptive to actual environment & Challenging optimization problem\\
       \hline
\end{tabular}
\end{table*}


\newcommand{\tabincell}[2]{\begin{tabular}{@{}#1@{}}#2\end{tabular}}
\begin{table*}[t]\centering
\renewcommand{\arraystretch}{1.6}
\caption{Comparison of existing wireless systems for supporting IoE \cite{JR:Xiaoming_JSAC_massive_access,book:Xiaoming}.}
\label{Table:massive_access_technlogies}
\begin{tabular}{|l|c|c|c|c|c|}\hline
  & Zigbee & Bluetooth & WiFi & LoRa & Cellular \\

\hline Spectrum & \tabincell{l}{Unlicensed} & \tabincell{l}{Unlicensed} & \tabincell{l}{Unlicensed} & \tabincell{l}{Unlicensed} & \tabincell{l}{Licensed}\\
\hline Connectivity & \tabincell{l}{Moderate} & \tabincell{l}{Small} & \tabincell{l}{Large} & \tabincell{l}{Massive} & \tabincell{l}{Massive}\\
\hline Throughput & \tabincell{l}{ Moderate}  & \tabincell{l} {Low} & \tabincell{l} {High} & \tabincell{l} {High} & \tabincell{l} {High}\\
\hline Coverage & \tabincell{l} {Short} & \tabincell{l} {Short} & \tabincell{l} {Moderate} & \tabincell{l} {Long} & \tabincell{l} {Long}\\
\hline Security level & \tabincell{l} {Moderate} & \tabincell{l} {Low} & \tabincell{l} {Moderate} & \tabincell{l} {High} & \tabincell{l} {High}\\
\hline Power consumption & \tabincell{l} {Low} & \tabincell{l} {Low}  & \tabincell{l} {High} & \tabincell{l} {Low} & \tabincell{l} {Low}\\
\hline Mobility & \tabincell{l} {No} & \tabincell{l} {No}  & \tabincell{l} {No}  & \tabincell{l} {Yes} & \tabincell{l} {Yes}\\
\hline Cost & \tabincell{l} {Low} & \tabincell{l} {Low}  & \tabincell{l} {Low} & \tabincell{l} {Moderate} & \tabincell{l} {High}\\
\hline
\end{tabular}
\end{table*}
\subsection{Future Research}

\subsubsection{Multiple Access}
A critical issue in realizing URLLC in UAV systems is the limited spectrum available for simultaneous ground-to-ground, ground-to-air, and air-to-air communications.
 Unlike terrestrial communication systems where ground BSs are connected to a data hub via high-speed fixed-line backhaul links, e.g., optical fibres, the implementation of ground-to-air and air-to-ground links usually relies on dedicated wireless communication channels. In particular, UAV-based communication systems have a stringent demand for system resources in both the ground-to-air and air-to-ground links due to the required support of high data rate backhauling and exchange of time-critical UAV control signals.  In this context, the allocation of  dedicated orthogonal radio resources to ground-to-air links is commonly considered in the literature. Yet, the spectral efficiency of orthogonal resource allocation is low. In fact, the wireless spectrum  is scarce and is already congested by existing communication systems.  Furthermore, in order to guarantee URLLC, multiple UAVs should be deployed simultaneously which puts a significant burden on the need for available spectrum.  Hence, conventional multiple access schemes based on orthogonal spectrum partition would quickly exhaust the available resources even for moderate numbers of UAVs and ground users. More importantly, this would introduce a long delay in the provisioning of URLLC and raises serious safety issues in controlling aerial nodes. Thus, efficient multiple access schemes based on non-orthogonal spectrum utilizations have to be developed for enabling  UAV-based URLLC systems.

\subsubsection{Hardware Imperfections}
Besides, existing algorithms for URLLC in UAV-based communication systems have been developed based on the assumption of perfect hardware. Yet, various types of hardware imperfections exist in practical UAV systems and must be accounted for \cite{JR:Vincent_HWI}.  For example, due to the finite precision of electronic circuits and the imperfection in manufacturing mechanical components,   the control of the UAV's trajectory and positioning may not perform as expected \cite{xu2020multiuser}.  As for the hardware modules for wireless communication, phenomena such as  power amplifier non-linearity, non-linear phase noise, frequency and phase offsets, in-phase and quadrature (I/Q) imbalance, and quantization noise jointly degrade the decoding capability of the receivers. Hence, it is still unclear whether it will be possible to guarantee URLLC in UAV-based communication systems. Thus,  a thorough performance analysis taking into account the impact of hardware impairments is needed.

\section{Massive Machine-type Communications} 

 The connection density of MTC networks will be around $10$ million devices per $\mathrm{km}^2$ by 2030 \cite{CISO:massive_access}.   Currently, for fast and cheap implementation, machine-type devices (MTD) access wireless networks via low-cost commercial technologies, e.g., Zigbee, Bluetooth, and WiFi. Besides,
the application of long range radio (LoRa) and cellular networks to the IoE has been proposed for enabling massive access. Indeed, it is  beneficial and economical to deploy the IoE using existing networks. However, traditional communication networks are designed and used for human-to-human (H2H) communications to serve a relatively small number of users, compared to the large number of devices in IoE, cf. Table \ref{Table:massive_access_technlogies}. 
  The fundamental challenges in the implementation of MTC are massive connectivity, ultra-high reliability and low latency requirements, and energy-efficient transmission. As  URLLC has been discussed in Section III, in this section, we tackle the challenges in MMTC from the perspectives of massive access and energy-efficient transmission.

  Unlike traditional communication networks which focus on the performance of the downlink for a small number of users, MTC  has a much higher demand on system resources in the uplink due to the associated massive uplink connectivity \cite{Magazine:MTC_UAV}. For instance, security cameras can be installed on UAVs for capturing images of  specific areas for security surveillance. The captured images are periodically conveyed to a ground BS for further data analysis. In this context, allocating dedicated radio resources to MTD orthogonally has been proposed in the literature, e.g., \cite{Magazine:MTC}, due to the simple design. However, for MTC, since the number of MTDs is large, orthogonal multiple access (OMA) would quickly exhaust the valuable radio resources and introduce exceedingly long delays. Furthermore, some commonly adopted random access protocols for H2H communications such as Carrier Sense Multiple Access (CSMA) and ALOHA are strictly suboptimal for MTC, since they suffer from congestion and overloading in the presence of a huge number of MTDs. Therefore, efficient multiple access mechanisms such as grant-free random access and NOMA \cite{Magazine:Liang_Lu_massive_access,JR:NOMA_Kwan_5,JR:NOMA_Kwan_4}, are  necessary to enable the massive connections required for MTC  communications. In addition, MTDs for IoE are usually powered by batteries with limited capacities in practice. Although the lifetime of the traditional H2H communication networks can be extended by replacing or recharging the batteries, this may be inconvenient, dangerous (e.g. in a toxic environment), and costly in MTC networks due to the large number of MTDs.  Generally speaking, the energy consumption challenge  to achieve green and self-sustainable  MTC  networks could be addressed by a proper combination of energy harvesting and/or energy-efficient designs which has drawn significant interest from both academia and industry \cite{JR:EE_massive_MIMO_Kwan,JR:QQ_EE_WPCN,COML:EE_WIPT,JR:energy_efficient_1,wu2016overview,zhang2016fundamental}.

The emergence of UAVs as a viable solution to deliver communication services in the past decades has introduced a paradigm shift in MTC.  Conventionally, long-distance communication between ground MTDs and their home BSs creates a system performance bottleneck which results in a strict limit on the lifetime of the communication network. In fact, by exploiting the flexility and high-mobility of UAVs, one can deploy a UAV as a sink node to collect data directly from ground MTDs. For example, for the case of a large sensor network deployed  in a rural area with limited cellular coverage,  a UAV can serve as a sink node to collect information data from the sensors \cite{918}. In particular, by exploiting the LoS communication channels from the UAV to the ground sensors, the sensors can transmit with low power which can help extend the lifetime of the sensing network. Furthermore, by taking into account  the periodic and sparse transmission characteristics of sensors, UAV semi-persistent scheduling can be performed to reduce control plane traffic between the UAV and the BS. On the other hand, as an MTD itself, a UAV can help relay the data from the ground MTDs so as to improve the energy efficiency of the whole communication system. However, despite the various practical advantages of deploying UAVs, the performance of UAV-enabled MTC is still limited by the small onboard battery equipped at UAVs due to the  size, weight, and power (SWaP) constraints of UAVs. Hence, practical methods for enabling sustainable MTC via UAV communications are needed.

\subsection{NOMA}

NOMA is a key enabling technology for future mMTC applications due to its capability of allowing simultaneous transmissions of multiple devices in the same resource block \cite{ding2017survey,book:Kwan_5G,JR:Xiaoming_JSAC_massive_access}.  Specifically, NOMA leverages superposition coding (SC) at the transmitters and successive interference cancellation (SIC) at the receivers to achieve efficient access as well as to partially mitigate co-channel interference. It has been shown that NOMA is particularly effective for cases where the users experience  substantially different channel conditions \cite{ding2017survey}.

\subsubsection{NOMA for Cellular-connected UAVs}
Due to its advantages, NOMA is practically appealing for application in cellular networks with co-existing UAV and ground users. Particularly, NOMA allows UAVs/a UAV swarm to reuse the resource blocks that are already occupied by ground users, thus improving the number of aerial users that can be supported even for a high ground user density, as compared to the non-scalable OMA scheme.
On the one hand, more reliable air-to-ground communications can be established due to the existence of LoS links, as compared to the non-LoS terrestrial channels between the ground users and BSs. Besides, the strong LoS air-ground channels allow the UAVs to be visible to a large number of ground BSs over a wide area. As a result, the UAVs can be potentially served by more BSs simultaneously than the ground users, thus achieving a higher macro-diversity gain in terms of user association. Besides, using NOMA in such unique air-ground channel environments also brings challenges since the UAVs' uplink transmissions will greatly degrade the received ground users' signals at a large number of ground BSs, considering the high frequency reuse factors of current multi-cell systems. Furthermore, ensuring that each individual BS can cancel the UAV's uplink interference via signal decoding may severely limit the performance gain of NOMA over OMA, since the UAV's achievable rate will be then limited by the BS with the worst channel conditions.  To address this issue, a new decode-and-forward (DF) based cooperative NOMA scheme, was proposed in \cite{mei2019uplink}, which exploits  interference cancellation among collaborative adjacent BSs with backhaul links, e.g., the existing X2 interface in LTE \cite{dahlman20134g}. Specifically,  some BSs with better channel conditions are selected to decode the UAV's signals first, and then forward the decoded signals to their backhaul-connected BSs for interference cancellation. It was shown that the proposed scheme achieves higher data rates than both OMA and  non-cooperative schemes, especially when the ground traffic is congested. To further improve the performance, a quantize-and-forward (QF) based cooperative interference cancellation approach was proposed in \cite{mei2020uplink} where the adjacent BSs only quantize the received UAV signals without decoding them.

In contrast, in the downlink, UAV receivers suffer from strong co-channel interference from a large number of ground BSs. However, the interference mitigation techniques proposed for uplink transmission are not applicable in the downlink because the roles of the UAVs have changed from interference sources to interference victims. One straightforward approach is to increase the transmit power of all ground BS. However, this provides only a marginal gain as the interference increases as well. Another possible practical approach is to leverage cooperative beamforming, where the available BSs that are not serving ground users in the UAV's occupied resource block, transmit collaboratively to the UAV to improve its received power to overcome the co-channel interference. However, as the ground user density increases, the number of such BSs drops quickly and the co-channel interference also increases, thus rendering this approach ineffective. To overcome this difficulty, a new cooperative beamforming scheme with interference transmission and cancellation (ITC) was proposed in \cite{mei2019cooperative}. Specifically,  the signals for  terrestrial users reusing the same resource block as the UAV are first forwarded by their home BSs to the BSs serving the UAV, and then transmitted along with the UAV's signals via cooperative beamforming. As a result, the desired signal power at the UAV receiver is improved and the terrestrial interference is also suppressed, without affecting the existing transmissions.   This centralized implementation requires excessive backhaul transmissions among different BSs, which may be costly to implement. To lower the implementation complexity and signaling overhead, a distributed algorithm requiring only local information exchange among BSs was proposed based on the concept of  divide-and-conquer in \cite{mei2019cooperative}. It was demonstrated that such a distributed design can still significantly improve the UAV's performance as compared to the conventional schemes without  ITC, especially when the terrestrial user density is high. However, current research results only consider a limited number of UAVs and the support of massive UAVs or UAV swarms is still a challenging problem that needs further efforts \cite{new2020robust}.

\subsubsection{UAV-enabled NOMA}
By proactively leveraging their high mobility, UAVs as aerial BSs inherently possess the ability to effectively exploit asymmetric channel conditions of different ground devices to realize the performance gains promised by NOMA. {A comparison of existing works is shown in Table \ref{tab:URRLC_techniques}.}
For instance,  the authors in \cite{wu2018capacity} studied UAV enabled NOMA transmissions to two quasi-static ground users. The capacity region was characterized by jointly optimizing the UAV’'s trajectory and transmit power/rate allocations over time, subject to the constrained maximum speed and transmit power of practical UAVs. A fundamental result is that regardless of the multiple access scheme, the optimal UAV trajectory follows a simple hover-fly-hover (HFH) policy. In particular, it was revealed that in the cases of  infinitely high UAV speed and/or long flight duration, NOMA and  time division multiple
access (TDMA) achieve the same rate performance. However, NOMA generally outperforms OMA (including TDMA and  frequency division multiple access (FDMA)), and  the capacity gain achieved by NOMA over  OMA decreases as the UAV maximum speed and/or flight duration increases.
The comparison of two-user NOMA and OMA for UAV-assisted communication was also extended to other design objectives such as sum-rate \cite{sohail2018non} and outage probability  \cite{sharma2017uav}.
In particular, \cite{sharma2017uav} considered a more practical Rician air-ground channel setup where the UAV flies with a constant speed following a circular trajectory, with the objective of minimizing the outage probability. The condition under which NOMA outperforms TDMA was also derived in terms of the channel and UAV trajectory parameters.

For a general setup with multiple users, proper user pairing with bandwidth allocation is also important to unlock the full potential of NOMA. In \cite{nasir2019uav}, the UAV was assumed to pair one near user (cell-centered) with one far user (cell-edge). Then, the multi-user rate max-min optimization problem was formulated by jointly optimizing the bandwidth allocation, power allocation, UAV altitude, and antenna beamwidth. {In subsequent works, the use of NOMA was extended to multiple-antenna and multi-UAV networks \cite{hou2018multiple,hou2019exploiting,8663350,duan2019resource,rupasinghe2019angle}. In \cite{hou2018multiple}, the downlink transmission from a multi-antenna UAV to multiple ground user clusters was considered where analytical expressions for the outage probability and the ergodic rate were derived. This was then extended in \cite{hou2019exploiting} by considering the use of multiple UAVs in a large-scale cellular network. In \cite{duan2019resource}, the user angle was exploited as feedback information for mmWave NOMA communications. Compared to the conventional limited feedback scheme based on users' distances, angle information was shown to have a significant potential in providing better separation for NOMA users in the power domain specifically for scenarios with multi-antenna transmission. In \cite{rupasinghe2019angle}, the resource allocation problem in a multi-UAV aided IoT NOMA uplink transmission system was studied where the channel assignment, the uplink transmit power, and the flying heights of UAVs were jointly optimized to maximize the system capacity.}

\begin{table*}[]
 \centering
\renewcommand{\arraystretch}{1.6}
\caption{Comparison of existing works on NOMA and UAV. }\label{tab:NOMA:UAV}
\begin{tabular}{|c|l|l|l|l|}
\hline
\textbf{Reference} & \textbf{System Setup}         & \textbf{Air-Ground Channel Model} & \textbf{Design Objective}                                                  & \textbf{Approach}        \\ \hline
\cite{wu2018capacity}           & Single UAV, two users         & LoS channel                       & Capacity region                                                            & Closed-form expression   \\ \hline
\cite{sohail2018non}         & Single UAV, two users         & Probabilistic LoS channel         & Sum rate maximization                                                      & Gradient descent         \\ \hline
\cite{sharma2017uav}         & Single UAV, two users         & Rician channel                    & \begin{tabular}[c]{@{}l@{}}Outage probability \\ minimization\end{tabular} & Closed-form expression   \\ \hline
\cite{nasir2019uav}            & Single UAV, multiple users    & LoS channel                       & Minimum rate maximization                                                  & SCA                      \\ \hline
\cite{8663350}           & Single UAV, multiple users    & LoS channel                       & Sum rate maximization                                                      & Two-step method          \\ \hline
\cite{duan2019resource}         & Multiple UAVs, multiple users & LoS channel                       & Sum rate maximization                                                      & SCA and Lagrange duality \\ \hline
\end{tabular}
\end{table*}

\subsection{Energy Harvesting and Energy-efficient Designs}

Besides multiple access schemes, promising  techniques to prolong the lifetime of MTDs have also been proposed including energy conservation schemes and wireless power transfer (WPT) techniques \cite{bin2011survey,QQWU_2016_txrx,2016Energy,zhang2018wireless,wu2018spectral,wu2015wireless,wu2019generalized}. Particularly, WPT provides several  competitive advantages over the other energy harvesting techniques such as its continuous availability and long service range.  However, the performance of energy harvesting at the MTDs critically depends on the locations of the power sources (e.g., BS, WiFi, etc.). For  MTDs that are close the power sources, a sufficient amount of energy can be harvested, conversely, for those which are far away from the power sources, energy harvesting may not be possible at all due to the high sensitivity of  typical energy receivers  \cite{wu2016overview,qing16_wpcn_twc}. Fortunately, UAVs with their high mobility provide a new promising solution to tackle these shortcoming of WPT. The first relevant work  \cite{2018UAVxu} studied a UAV-mounted energy transmitter  broadcasting wireless energy to charge  two MTDs. The results showed that  when the distance between the two MTDs is very small,  the Pareto boundary of the energy region can be  achieved by fixing the UAV's horizontal position. In contrast, when the distance between the two MTDs is large, to achieve the Pareto boundary of the energy region, the UAV should fly and hover between the two MTDs.  Subsequently, this work was extended to a more general setup with multiple MTDs  in \cite{xu2018uavenabled} where the authors studied two different objective functions, namely the  maximization of the sum energy harvested by  all the MTDs and  the max-min energy harvested  among  MTDs. Different from the above works, which obtained locally optimal solutions, a globally optimal one-dimension UAV trajectory design was developed with the aim of maximizing the minimum energy received by either one of  the MTDs \cite{hu2019optimal}.

 In addition, UAV-aided wireless powered communication network (WPCN) is drawing considerably attention  in the research community \cite{zeng2019accessing}. Specifically, receiver devices in WPCN first harvest energy from the signals sent by the UAV in the downlink, and then utilize
the harvested energy to transmit information to the UAV in the uplink. Depending on the configuration of  the UAVs,  the UAV-aided WPCN  can be   classified into two categories. {The first is  the  \textit{integrated UAV-aided  WPCN}, where the energy transmitter and information receiver are carried  by one UAV \cite{yang2020enregy,xie2019throughput,xie2020common,tang2020minimum}.
The other is the\textit{ separated  UAV-aided WPCN}, where the energy transmitter and information receiver are  installed on  two different UAVs \cite{park2019UAV,wu2019Minimum}.  Specifically, the system energy maximization problem and the max-min throughput optimization problem in UAV-enabled WPCNs were studied in \cite{yang2020enregy} and \cite{xie2019throughput}, respectively. Then, \cite{xie2020common} extends the work in  \cite{xie2019throughput} to a two-user interference channel for WPCN employing two UAVs. In \cite{tang2020minimum}, the multi-agent deep reinforcement learning based algorithm was proposed to tackle the max-min optimization problem in a general multi-UAV enabled WPCN.
Furthermore,  UAV-enabled simultaneous wireless information and power transfer (SWIPT) was proposed recently, where the UAV is  deployed as an aerial BS to simultaneously send
information and energy to the receivers on the ground \cite{huang2019UAV,kang2020joint,feng2020uav}.  Specifically, the  UAV's trajectory, transmit power, and the users' power splitting ratios were jointly optimized to maximize the minimum achievable rate among the users with their harvested energy constraints, whereas in \cite{huang2019UAV}, the max-min harvested energy among users was considered subject to users' achievable rate constraints. An overview on UAV-enabled SWIPT was presented in \cite{feng2020uav} by considering the context of IoT works for emergency communications.}

On the other hand, different from terrestrial BSs which are connected to the power grid,  UAVs are typically powered by  on-board batteries with limited energy storage capacity, which thus drastically limits their  endurance. To overcome the short flight time  of UAV, several techniques  have been proposed, which can be classified into two categories: external-powered UAV and mechanical dynamic optimization.
\subsubsection{External-powered UAV}
For external-powered UAVs, depending on the external sources used for powering the UAV, they can be further  divided into two approaches, namely
solar-powered and laser-powered UAVs. For solar-powered UAVs, a proper size solar panel is installed on the UAV,  which first harvests solar energy  and then  converts it to electrical energy enabling the UAV to fly and communicate  \cite{morton2015powered,Oettershagen2016Perpetual,9201322}. Solar-powered UAVs were experimentally verified  in  \cite{morton2015powered,Oettershagen2016Perpetual}. The results revealed that by carrying a solar panel, the UAV can continuously fly for more than $24$ hours, which renders this approach appealing in practice. However, the amount of harvested solar energy depends heavily on the flight  altitude, weather conditions, and temperature  \cite{Duffie1991Solar,sun2019Optimal}. In particular, it was shown in \cite{sun2019Optimal}  that there exists a tradeoff between the  harvested solar energy and  the quality-of-service (QoS) for the ground users.  Specifically, if the UAV operates above the clouds, it can harvest more solar energy but incurs high path loss between the UAV and the ground users due to the high flight altitude. If the UAV flies below the clouds,  less solar energy is available since the solar energy is mainly absorbed by the clouds, while the quality of the communication link between the UAV and the ground users is improved.  {Instead of exploiting solar energy, a photovoltaic receiver can be installed on the UAV to harvest power from a laser power beacon. Compared to solar power, the laser-beam power supply is more stable and can deliver much more energy to the receiver with a narrowly focused energy beam \cite{Ouyang2018throughput,zhao2020efficiency,8866716,9203992}. In particular, a laser-powered UAV
wireless communication system was studied in  \cite{Ouyang2018throughput}, where a laser transmitter sends laser beams to charge a fixed-wing UAV in flight, and the UAV uses the harvested laser energy to communicate with a ground station.
In  \cite{zhao2020efficiency},  the weighted information transmission efficiency  and the laser power transmission efficiency of a rotary-wing UAV enabled relay system were maximized subject to the information/energy-causality constraints, the power budget constraints and the UAV’s mobility constraints. It was shown that the UAV's trajectory is highly related to the laser wavelength and the weather condition.
}

\subsubsection{ Mechanical dynamic optimization}
For mechanical dynamic optimization, the research was mainly focused on how to minimize the UAV's communication energy and propulsion energy consumption \cite{zeng2017energy,980,Mozaffari2016Optimal,Mozaffari2016mobile,Kaleem2019UAV,hua2018power,hua20203D,wuyang2020energy,duo2020energy}.
The UAV's communication energy consumption comprises the circuit power consumption as well as the transmit power consumption. Some relevant works optimized the deployment of the UAV to minimize the transmit power while achieving the different design objectives such as the users' QoS, communication coverage range, and delay \cite{Mozaffari2016Optimal,Mozaffari2016mobile,Kaleem2019UAV}. However, accurately modelling the propulsion energy consumption is generally difficult since it crucially depends on many practical factors, including the UAV velocity, the UAV acceleration, etc. Accordingly, in \cite{zeng2017energy}, a theoretical propulsion energy consumption  model for fixed-wing UAVs was derived, which is given by
\begin{align}\small
\!\!\!P \!=\! \left| {{c_1}{{\left\| {\bf{v}} \right\|}^3} + \frac{{{c_2}}}{{\left\| {\bf{v}} \right\|}}\left( {1 + \frac{{{{\left\| {\bf{a}} \right\|}^2} - \frac{{\left( {{{\bf{a}}^T}{\bf{v}}} \right)}}{{{{\left\| {\bf{v}} \right\|}^2}}}}}{{{g^2}}}} \right) + m{{\bf{a}}^T}{\bf{v}}} \right|, \label{UAV_propulsion_energy}
\end{align}
where $c_1$ and  $c_2$ are constants which are related to the air density, drag coefficient, and wing area; $g$ and $m$ represent the gravitational acceleration and  the aircraft's mass, respectively; ${\bf{v}}$ and ${\bf{a}}$ denote the UAV's velocity and acceleration vectors, respectively. It is observed  from \eqref{UAV_propulsion_energy} that when the UAV velocity is equal to ${\bf{0}}$, the UAV's propulsion energy tends to  infinity, which indicates that the UAV trajectory should be carefully designed to avoid excessive energy consumption and improve endurance. However, for communication service provisioning, it is desirable to deploy the UAV hovering above the ground users \cite{hua2018power,JR:wu2018joint}. As a result, there is a tradeoff between minimizing the  UAV's energy consumption and maximizing the communication throughput.

To address this tradeoff, a new metric, namely the energy efficiency defined as the ratio of the throughput and the  UAV energy consumption, has been proposed and widely adopted to strike a balance between the above two important design goals \cite{zeng2019accessing}. {Different from the notion of energy efficiency adopted  in  traditional cellular networks, where the energy consumption is related to the transceiver's circuit power consumption and transmit power, energy-efficient designs for UAVs are also subject to the UAV's dynamic and kinematic/mobility constraints, which makes the design problems highly non-convex and much more challenging. For example, the UAVs' trajectories are continuous and need to satisfy a certain velocity, acceleration and turning angle constraints.
Some commonly used methods to solve the resulting optimization problems including successive convex approximation (SCA), Majorization-Minimization (MM) method, block coordinate descent method (BCD),  alternative direction multiplier method (ADMM), penalty based method, machine learning, etc.}
A handful of works have considered  energy-efficient UAV design  for  different applications such as UAV-assisted mobile edge computing and UAV-assisted backscatter communication \cite{hua2019energy,duo2020energy,li2020energy,yang2019energy}.

{However, current research results only focus on a single UAV or few UAVs,  while energy-efficient design for a large number of collaborative UAVs is still a key challenge due to the excessive  data  and control  information exchange between them needed for efficient cooperation. To tackle this challenge, some heuristic methods can be applied to the UAV trajectory optimization, such as particle swarm optimization (PSO) as well as its several evolutionary paradigms \cite{kennedy1995particle,kim2015trajectory,roberge2012comparison,foo2006three}, derivative-free stochastic optimization and Hamiltonian optimization \cite{liu2018age}. Whereas those algorithms are generally suboptimal and are not guaranteed to obtain the globally optimal solution, the genetic algorithm (GA) is a useful tool to find optimal solutions for some optimization objectives such as energy consumption \cite{gemeinder2003ga,roberge2012comparison}. However, it leads to a relatively high computational complexity and associated delay in practice, especially when the number of UAVs is large.
Recently,  game theory-based UAV control methods, e.g. mean-field game,  have been exploited as efficient methods to model the interactions among a large number of UAVs \cite{9290094,9309341,zhang2020trajectory,zhang2019cooperation}. For example,  the mean-field approximation was  leveraged in \cite{zhang2020trajectory} to obtain the near optimal trajectory in the context of UAV-to-device underlaid cellular networks and the approximation becomes more accurate as the number of involved UAVs becomes larger.  Furthermore, the joint trajectory design and radio resource management problem for cellular Internet of UAVs has been studied in \cite{zhang2019cooperation} by leveraging a cooperative sense-and-send protocol.}

\subsection{Future Research}

\subsubsection{Air-ground Energy Consumption Tradeoff}
For many practical data collection applications in mMTC networks, a large amount of  devices may be distributed in a wide area while each of them only has small bursts of data to transmit. In this case, having UAVs  getting close to each of them to collect data may lead to high propulsion energy consumption. This thus leads to an energy consumption tradeoff between UAVs and ground devices \cite{yang2018energy}. To address this issue, inter-UAV cooperation among UAV swarms is a promising solution where a UAV cluster head can be dynamically selected and other UAVs first collect data in a small area and then transmit to the UAV cluster for further processing.

\subsubsection{Access Protocols and Security}
{For the case of large-scale deployment of UAVs in heterogenous applications, the communication activities of UAVs to the same ground BS can be highly random, which may call for  random access protocols. Although the  grant-based random access protocol is simpler in practical implementation, as the number of UAVs grows quickly,  it would result in a high probability of access failures due to collision as well as high signaling overhead arising from handshakes. As a consequence, grant-free random access or unsourced random access protocols with properly designed UAV grouping methods help achieve efficient UAV swarm communications \cite{JR:Xiaoming_JSAC_massive_access}. In \cite{ding2019simple}, a semi-grant-free transmission based on NOMA was proposed to offer more refined admission control while minimizing the system signalling overhead.  However, a fair and complete comparison between the grant-free random access against NOMA for 3D wireless networks with UAVs remains still an open problem that merits further investigation \cite{choi2020grant}.  } Moreover, security and privacy are other important issues in safeguarding wireless communication with massive UAV and ground devices \cite{wu2019safeguarding,zhang2018securing,li2018uav,cui2018robust,Xiaobo18uav,duo2020energy,xinrong2020UAV,li2018uav}.

\begin{figure*}
\centering
\includegraphics[width=0.8\textwidth]{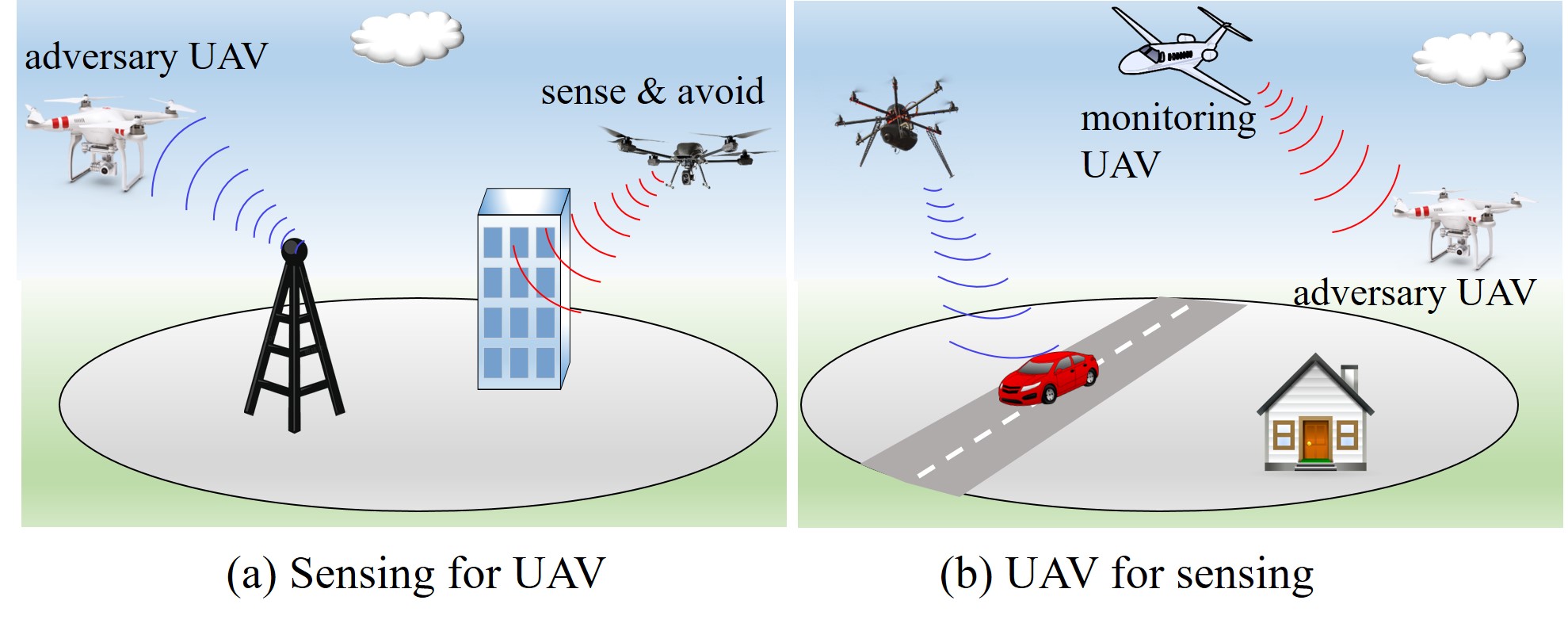}
\caption{{Two paradigms of sensing in UAV networks: (a) Sensing for UAV; (b) UAV for sensing.}}
\label{F:UAVSensing}
\end{figure*}

\section{{Radio-based Sensing}}
Besides the requirement of high-performance wireless communications, the ability to support effective and efficient sensing is also essential for realizing the vision of integrating UAVs into 5G-and-beyond networks. Commercial UAVs nowadays are already equipped with a multitude of sensors of various types, such as inertial measurement unit (IMU), accelerometers, tilt sensors, and current sensors. Such embedded sensors provide important real-time information for ensuring safe UAV operation, such as the UAV's position and orientation estimates, direction and flight path maintaining, and power consumption management.

 On the other hand, for future wireless networks, where UAVs of large-scale deployment will be seamlessly integrated into terrestrial communication systems, sensing merely relying on those on-board embedded sensors will be insufficient. Instead, a combination of both UAV embedded sensing and infrastructure-based sensing is needed to achieve high sensing performance, in terms of response time, sensing range, coverage, reliability, accuracy, and efficiency. In this section, we focus on  radio-based sensing, where the detection and parameter estimation of the targets of interest are based on the radio signals echoed/scattered  by the targets. Compared with alternative sensing technologies such as acoustic-based, vision-based, and light-based sensing, radio sensing is less vulnerable to poor environmental conditions (such as noisy background or dark areas) and can typically support a larger sensing range. Besides, the fact that both radio sensing and wireless communications utilize radio signals to accomplish their tasks makes it necessary to study both systems jointly.

As illustrated in Fig.~\ref{F:UAVSensing}, {radio-based UAV sensing} can be classified into two main paradigms, namely {\it sensing for UAV} and {\it UAV for sensing}. In the former case, sensing technologies  are utilized to support safe UAV flight and  low-altitude airspace monitoring and traffic management. By contrast, for the paradigm of {\it UAV for sensing}, dedicated UAVs are dispatched as aerial flying platforms to provide sensing support from the sky. {In addition,  a comparison of existing works is shown in Table \ref{Table:sensing}.}

\begin{table*}[]
\centering
\renewcommand{\arraystretch}{1.6}
{\caption{Comparison of existing works on radio-based UAV sensing.}\label{Table:sensing}
{\begin{tabular}{| p{2cm} | p{2cm} |p{9cm}|}
\hline
& \textbf{Reference} & \textbf{Main contributions} \\  
\hline
\multirow{4}{*}{\textbf{Sensing for UAV}}&
\cite{1225} &Overview of sense and avoid technologies for unmanned aircraft systems\\  
\cline{2-3}
& \cite{1218} & Key techniques of detection, tracking and interdiction of amateur UAVs \\ 
\cline{2-3}
& \cite{1221} & Classification of micro-drones into loaded or unloaded vehicles using multi-static radar \\
\cline{2-3}
& \cite{1223} &  Classifying UAVs versus other flying objects such as birds, by extracting the features of the tracks generated by radar \\
\hline

\hline
\multirow{5}{*}{\textbf{UAV for Sensing}}&
\cite{1216} &Overview for applications of UAV-based sensor platforms and experiments for imaging radar\\  
\cline{2-3}
& \cite{1226} & UAV-based remote sensing for field-based crop phenotyping \\ 
\cline{2-3}
& \cite{1224} & UAV-based synthetic aperture radar (SAR) to enhance the radar aperture for high angular resolution
 \\
\cline{2-3}
& \cite{1222} &  Multiple flying UAVs with optimized trajectories to localize moving RF sources \\
\cline{2-3}
& \cite{1217} & Dynamic and reconfigurable aerial radar network composed of UAVs to detect and track other unauthorized/malicious UAVs\\
\hline
\end{tabular}}}
\end{table*}

\subsection{Sensing for UAV}
Two typical use case scenarios of sensing for UAV are {\it sense-and-avoid} (SAA) and {\it adversary UAV detection, tracking, and classification}.

 SAA plays an indispensable role to ensure safe UAV flight, especially for autonomous or semi-autonomous UAVs \cite{1225}. Different from manned aircrafts, UAVs do not have a pilot onboard, and thus they  heavily rely on their own sensed information for swift response for collision and/or obstacle avoidance. Note  that even for UAVs under real-time remote control, SAA is important for avoiding catastrophic events, considering the long round-trip radio propagation delays and the extra response time of ground pilots. To achieve fast response  even in highly dynamic environments, SAA typically relies on the sensed information of on-board sensors, though ground based SAA (GBSAA) is also possible. Note that besides sensing for collision/obstacle avoidance, in many applications such as UAV-enabled fertilizer or pesticide spraying, sensing may also be necessary for constant-altitude maintenance so as to ensure uniform spraying \cite{1216}, which is a challenging task in tough terrains. While many contemporary commercial UAVs are already enabled with SAA capabilities, most of them are vision- or light-based\footnote{https://www.dronezon.com/learn-about-drones-quadcopters/ptop-drones-with-obstacle-detection-collision-avoidance-sensors-explained/}, which makers them vulnerable to poor environmental conditions.

 Another important use case of sensing for UAV arises from the imperative need for the detection, tracking, and classification of potentially illegitimate and hazardous UAVs. Besides  their benign usage, UAVs could also be misused, either intentionally or unintentionally, which may jeopardize public safety and/or threaten privacy. Under such circumstances, UAVs are usually non-cooperative or even deceptive, which renders applying active detection and localization methods infeasible \cite{1208}, but rather requires reliance on radar sensing based on passively echoed/scaterred signals. While radar sensing for aircraft detection has long been used in military, its application for UAVs is facing new challenges. For instance, different from manned aircrafts, UAVs usually have much smaller radar cross section (RCS), which makes them more difficult to detect. Besides, rotary-wing UAVs  are capable of flying at low speed or hovering, which makes them difficult to separate from stationary clutter background or other natural flying objects such as birds.

 Radar sensing for UAV networks has received significant research attention recently. For example, the key techniques of detection, tracking, and interdiction of amateur drones are discussed in \cite{1218}. Beyond UAV detection, UAV classification is also important for safety and privacy protection in the future internet-of-drones (IoD) era. For example, in \cite{1221}, the authors studied the classification of micro-drones into loaded or unloaded vehicles using multi-static radar. In \cite{1223}, the authors studied the problem of classifying UAVs versus other flying objects such as birds and manned aircrafts, by extracting the features of the tracks generated by radar,  where a track is defined as a series of plots associated with each target observed in consecutive radar scans.

\subsection{UAV for Sensing}
Another promising paradigm of UAV sensing is to utilize UAVs as aerial nodes to provide wireless sensing support from the sky, which we refer to as {\it UAV for sensing}. Compared with conventional ground sensing, UAV-based sensing has several advantages. First, thanks to its elevated altitude and reduced signal blockage, UAV-based sensing typically has a wider field of view (FoV) compared to ground sensors. Besides, the highly controllable 3D UAV mobility makes it possible to flexibly deploy UAV sensors to hard-to-reach areas, such as poisonous or hazardous locations. Furthermore, the high mobility of UAVs also introduces  a new design degree of freedom (DoF) for sensing performance optimization, via 3D sensor trajectory optimization. This is particularly appealing for target tracking, where the UAV locations can be dynamically adjusted to best track the target. Therefore, UAV-based sensing has a wide range of potential applications, such as law enforcement, precision agriculture, 3D environment map construction, search and rescue, and military operations.

 Due to the aforementioned advantages, UAV-based sensing has received growing interest recently. For instance, various applications of UAV-based sensor platforms were outlined in \cite{1216}, together with a discussion of  experiments with an imaging radar. In \cite{1226}, the authors surveyed UAV-based remote sensing for field-based crop phenotyping. In \cite{1224}, UAV-based synthetic aperture radar (SAR) was used to enhance the radar aperture for high angular resolution. Besides, multiple flying UAVs were utilized in \cite{1222} to localize moving RF sources, and the UAV trajectory was optimized to minimize a bound on the achievable localization error.  UAV-aided air quality sensing was discussed in \cite{1246}. In \cite{1217}, the authors proposed a dynamic and reconfigurable aerial radar network composed of UAVs to detect and track other unauthorized/malicious UAVs. It was demonstrated that by exploiting the new DoFs offered by UAV trajectory optimization, the dynamic aerial radar network is able to improve the tracking performance over a conventional terrestrial radar network with fixed deployment.

\subsection{System Model and Promising UAV Sensing Technologies}
For a basic UAV-based radar system with $N_T$ transmit antennas and $N_R$ receive antennas, the input-output relationship can be modelled as
\begin{align}
\mathbf y(t)=\sum_{k=1}^K \alpha_k \mathbf a(\theta_k^{(R)},\phi_k^{(R)})\mathbf b^T(\theta_k^{(T)},\phi_k^{(T)})\mathbf x(t-\tau_k)+\mathbf n(t),\label{eq:model}
\end{align}
where $K$ denotes the total number of targets, $\mathbf y(t)\in \mathbb{C}^{N_R\times 1}$ and $\mathbf x(t)\in \mathbb{C}^{N_T\times 1}$ represent the received and transmitted signal waves, respectively, $\alpha_k$ is the complex-valued coefficient from the transmitter to the receiver associated with the $k$th target, $\theta_k$ and $\phi_k$ are the elevation and azimuth angles of the $k$th target, respectively, with the superscripts $(\cdot)^{(R)}$ and $(\cdot)^{(T)}$ denoting the angle-of-arrival (AoA) and angle-of-departure (AoD), respectively. Note that different from ground sensing where the azimuth angle is usually of primary interest, for UAV-based  sensing, both the azimuth and elevation angles are important due to the elevated UAV position. Furthermore, $\mathbf a(\cdot)$ and $\mathbf b(\cdot)$ denote the receive and transmit array responses, respectively, $\tau_k$ is the radio propagation delay for signals reflected/scattered by the $k$th target, and $\mathbf n(t)\in \mathbb{C}^{N_R\times 1}$ denotes the receiver noise. Note that in \eqref{eq:model}, we have only included the signal components reflected/scattered by the $K$ targets of interest. This may be a valid model for UAV sensing in wide-open areas or after proper clutter subtraction, when the signal components originating from background clutter have been properly eliminated. The radar performance is critically affected by the radar waveform $\mathbf x(t)$, which is usually designed to have good autocorrelation properties. One popular waveform is the linear frequency-modulated continuous waveform (FMCW) \cite{1236}, where multiple chirp signals are transmitted with the frequency linearly increasing with time. Different from wireless communication systems, the transmitted radar waveform $\mathbf x(t)$ is known at both the transmitter and receiver. Therefore, based on the knowledge of the transmitted and received waveforms $\mathbf x(t)$ and $\mathbf y(t)$, the two fundamental problems in radar sensing are:

 {\bf Radar detection:} Detect the presence/absence of target $k$, i.e., $\alpha_k=0$ for hypothesis $\mathcal H_0$ and $\alpha_k\neq 0$ for hypothesis $\mathcal H_1$. In this case, the sensing performance can be measured in terms of the probability of correct detection/misdetection and the probability of false alarm.

 {\bf Radar estimation:} If target $k$ is present, estimate the key parameters $\theta_k^{(T)}, \phi_k^{(T)}, \theta_k^{(R)}, \phi_k^{(R)}$, and $\tau_k$, based on which the target state, such as location and moving velocity, can be further determined. Typical performance metrics for radar estimation include the accuracy or mean squared error (MSE) of the estimation. For unbiased estimators, the classic Cramer-Rao lower bound (CRLB) provides the theoretical performance bound on the variance of the estimate \cite{71}. In \cite{1239}, the authors proposed a new performance metric for radar estimation that is similar to communication rate, called radar estimation rate. Specifically, the target is treated as a passive node with some entropy about its own state that unwillingly communicates with the radar receiver, and radar estimation rate corresponds to how much additional information is gained about the target's state after the radar estimation process. Other sensible radar performance metrics include the number of resolvable targets, response time, and time and range resolution.


Note that \eqref{eq:model} is applicable for both bi-static and mono-static radars. In the former, the radar transmitter and receiver are geographically separated, whereas for the latter, they are collocated.  For mono-static radar, \eqref{eq:model} can be further simplified since we have $N_T=N_R$, $\theta_k^{(R)}=\theta_k^{(T)}$, $\phi_k^{(R)}=\phi_k^{(T)}$, and $\mathbf b(\cdot)=\mathbf a(\cdot)$. Besides, $\tau_k$ corresponds to the round-trip delay between the radar transmitter/receiver and the $k$th target, which is related to the target distance $r_k$ and radial velocity $v_k$ as $\tau_k=\frac{2(r_k+v_kt)}{c}$, where $c$ is the speed of light.

The technologies for radar sensing in general have been advanced tremendously during the past decades. In particular, it has been shown that MIMO radar is a powerful approach to enhance the radar performance. Specifically, for an $N_R\times N_T$ MIMO radar with orthogonal waveforms transmitted from the $N_T$ transmit antennas, we may achieve the same effective angular resolution as a virtual phased array radar with $N_TN_R$ elements \cite{1228,1227,1229}. This thus lays a strong foundation for cellular sensing to achieve super resolution with massive MIMO in 5G-and-beyond networks. Besides,  full-dimensional MIMO  radar also provides angular resolution in both the azimuth and elevation domains, which is essential for UAV sensing.

Another promising  technology for UAV sensing is mmWave sensing, which offers two promising advantages. First,  compared to sub-6 GHz systems, physical objects are electrically larger in mmWave systems because of their shorter wavelength. Thus, small objects like micro-UAVs that might be invisible for microwave systems become more visible and more easily detectable for mmWave systems \cite{1219}. To illustrate this fact, the authors in \cite{1219} used the radar cross-section (RCS) as a performance measure to compare the delectability of small objects with different frequencies. It was shown that a small drone illuminated by mmWave radar (60 GHz) has a RCS about 30 dB  higher than that by a 2 GHz microwave radar. Second, mmWave systems usually have a large bandwidth, which leads to high time and range resolution. In particular, it is known that the radar range resolution $\Delta$ is given by $\Delta=c/(2B)$, where $c$ is the speed of light and $B$ is the system bandwidth. This thus makes mmWave systems a promising technology for high-performance radar detection and estimation. Therefore, mmWave UAV  detection has received significant research interest, as reviewed in \cite{1219}.

\subsection{Joint UAV Communication and Sensing}
Traditionally, wireless communication and radar sensing were considered to be two independent systems that are fully separated in spectrum. With the continuous spectrum expansion of cellular communication systems (from sub-6 GHz to mmWave or even TeraHertz) and the need for developing more intelligent wireless networks, there has been a significant research interest recently in joint communication and  sensing (JCAS) \cite{1234,1235,1241,1242}. On the one hand, JCAS facilitates the design of a highly flexible and efficient systems utilizing the spectrum allocated for both purposes, which thus offers a new solution approach for resolving the spectrum scarcity. On the other hand, JCAS enables the sharing of wireless infrastructure and RF hardware, so as to build compact and light-weight wireless equipment with both communication and sensing capabilities. Note that this feature is particularly appealing for UAV networks, considering the severe SWaP and endurance constraints of UAVs, together with the significant role that radar sensing would play towards the integration of UAV into 5G-and-beyond cellular networks.

The preliminary level of JCAS is communication-radar coexistence \cite{1243}, where communication and radar sensing are still designed as two separate systems with their own respective waveforms and transmitter/receiver designs, but their mutual interference is properly controlled via resource allocation. This usually corresponds to the scenario when radar and communication share neither the transmitter nor the receiver. For example, a low-altitude UAV that continuously broadcasts radar waveforms for SAA needs to take into account the potential interference caused to the communication between nearby BSs and ground users, and vice versa.

A more advanced approach for JCAS is known as {\it dual function radar-communications} (DFRC) \cite{1245}, where both the radar and communication functions are implemented in the same device. DFRC is ideally suited for application in UAV networks. For example, a DFRC-enabled cellular BS may simultaneously communicate with the ground users it serves and monitor the low-altitude airspace to detect adversary UAVs. 
DFRC can be loosely classified as {\it radar-centric}, {\it communication-centric}, and {\it integrated design}. For radar-centric DFRC, the transmitted waveform $\mathbf x(t)$ in \eqref{eq:model} is radar-oriented, such as the traditional FMCW, but with additional information-bearing symbols embedded for communication, e.g., via phase or frequency modulation of the FMCW pulse \cite{1238}. For MIMO radar with transmit beamforming, the information-bearing symbols may also be embedded into the radar beam pattern, e.g., by amplitude modulation of the radar side-lobe levels \cite{1244}. While power-efficient and simple to implement, such radar-centric DFRC can only support very low data rates. On the other hand, for communication-centric DFRC, the system is primarily designed for wireless communications, with the additional radar functionality. For example, there has been significant interest in utilizing the standard  orthogonal frequency-division multiplexing (OFDM) communication waveforms for radar sensing \cite{1237,1230,1231}. By optimizing the various OFDM parameters, such as the sub-carrier spacing and power allocation, different trade-offs between communication and radar sensing performance can be achieved. However, for high mobility scenarios like UAV networks, the sensing performance (such as the unambiguous range) with OFDM waveforms is vulnerable to sub-carrier misalignment \cite{1232}. Lastly, for integrated radar-communication design, the DFRC waveform might be derived from the scratch without restriction to radar- or communication-centric designs, which makes  further performance improvements possible. For example, in \cite{1233},  the weighted combination of the radar and communication waveforms was used as the transmitted waveform of a DFRC MIMO transmitter, and the resulting transmit covariance matrix was optimized to preserve the desired radar beampattern, while guaranteeing a minimum required SINR requirement at each communication user. Beamforming for JCAS with analog or hybrid analog/digital antenna architectures was studied in \cite{1240} and \cite{1234}, respecitvely.

\subsection{Future Research}
The research on sensing in UAV networks is still in its infancy, with many new opportunities and challenges ahead. For example, due to the severe SWaP constraint, more compact, light-weight, and energy-efficient sensing devices are required for UAV-based sensing than for conventional ground-based sensing. Besides, the limited endurance of UAVs makes it quite challenging to build an aerial sensing network merely relying on UAVs that is available all day. In this regard, future sensing networks are expected to be highly heterogeneous, and may utilize  complementary ground, aerial, and even space sensing platforms to maximize sensing performance, which deserves in-depth investigations.  On the other hand, while JCAS has been studied extensively for terrestrial systems, e.g., for automotive applications, its application in  UAV networks is still in an early stage. In particular, with elevated UAV positions, the coverage requirement for both communication and sensing extends from the traditional 2D plane to the 3D  airspace, for which cost-effective networking architectures and signal transmission/reception techniques need to be developed. Besides, when UAVs are utilized as aerial communication and sensing platforms, the new DoFs offered by UAV trajectory design should be exploited for improving the sensing and communication performance. To resolve the SWaP problem of UAVs, energy-efficient designs for JCAS in UAV networks, which take  into account the UAV propulsion energy as well as the communication and sensing energy, deserve further investigation. Another promising research direction is to exploit the sensed information, such as the 3D radio propagation environment, to  enhance UAV communications \cite{zeng2020towards}.

\section{AI Integration}

In addition to supporting communication and sensing services, the integration of AI is expected to be another important feature of future cellular networks, towards the vision of network intelligence  \cite{Letaief_AI_Wireless2019}. Generally speaking, the interplay between AI and wireless networks has enabled two emerging paradigms, namely, AI-empowered wireless communications (see, e.g., \cite{Letaief_AI_Wireless2019,Gunduz_JSAC_2019}) and edge intelligence (see e.g., \cite{ZhouZhiandChenxu_Edge_Intelligence_2019,Guangxu_edge_lerning2020}).  In the former paradigm, AI and machine learning techniques are utilized as a new data-driven mathematical tool (in contrast to conventional model-driven methods) for optimization of wireless systems to enhance the communication performance. In the latter paradigm, AI and mobile edge computing (MEC) capabilities \cite{Mao_MEC_2017,WangXu_WirelessPoweredMEC_2018} are incorporated into BSs and APs at the network edge, for enabling various intelligent applications (such as autonomous driving and industrial automation) with extensive communication and computation requirements \cite{ZhouZhiandChenxu_Edge_Intelligence_2019,Guangxu_edge_lerning2020}. As compared to the conventional cloud and on-device intelligence, edge intelligence can significantly reduce the end-to-end latency and minimize the traffic loads to the core network.

Following the above two paradigms, AI is expected to also play an important role for 5G-and-beyond UAV communication networks in two aspects. On the one hand, for highly dynamic UAV-enabled 3D networks, AI and machine learning can serve as promising alternative mathematical tools to solve, e.g., joint UAV trajectory and communication design problems, where the conventional model-driven optimization may not work well due to the difficulty in obtaining accurate network state information. On the other hand, the integration of UAVs in edge intelligence not only facilitates emerging applications (such as drone VR and drone swarms), but also introduces new challenges in effectively handling computation-intensive and latency-critical AI tasks from the sky. This calls for the joint design of the UAVs' mobility/trajectory control together with the communication and computation resource allocation, which, however, may be particularly difficult as UAVs may act as aerial users or aerial edge servers or both.


{{Notice that although there have been prior works \cite{AIIntegration2020A,AIIntegration2020B} that reviewed the design of UAV communication networks based on machine learning, there still lacks an overview of the integration of AI in UAV communication networks from the above two aspects.}} In this section, we first review machine learning methods for UAV trajectory and communications design, then discuss  computation offloading design for UAVs with MEC, and finally present distributed edge machine learning with UAVs, followed by some open research problems.


\subsection{Machine Learning for UAV Trajectory and Communication Design}

The joint design of the UAVs' movements over time (e.g., deployment locations and flight trajectories) and the communication resource allocation is crucial for optimizing the performance of 5G-and-beyond UAV communication networks. Conventionally, such joint design is implemented in an offline manner based on model-driven optimization approaches, where the network state information (such as the locations of the communication nodes, the CSI, and the service request information) is assumed to be (perfectly or partially) known prior to the optimization. In this case, the joint design can be formulated as a deterministic optimization problem that is generally solvable via convex and non-convex optimization techniques. Such offline designs, however, may not work well in practice, due to the time- and spatial-varying traffic demands, the user mobility, and the complicated channel propagation environments introduced by the UAVs. To tackle this challenge, researchers have shifted from conventional model-driven approaches to alternative data-driven approaches by exploiting emerging machine learning techniques.

In general, machine learning can be classified into three categories, namely supervised learning, unsupervised learning, and RL. RL is particularly useful for the joint UAV movement and communication design, {{and thus will be the main focus of this subsection}}.\footnote{{{Please refer to \cite{AIIntegration2020B} for reviews of using other machine learning methods in the design of UAV communication networks.}}} To be specific,  RL optimizes the actions of one or more agents  in an environment to maximize the cumulative reward over a certain time horizon \cite{book_Reinforcement_Learning}. Supposing that UAVs are the agents of interest, then RL can be efficiently utilized in  UAV communication networks for rapidly adapting to the dynamic environment \cite{XLiu2020Ar_AIforUAV}, by properly choosing the UAVs' actions (e.g., deployment/trajectory design and resource allocations) and the reward functions (e.g., communication rate). In the literature, there are two lines of research that exploit  RL for optimizing the UAVs' operation in UAV-assisted communication systems \cite{XLiu2020Ar_AIforUAV,Bayerlein_UAV_RL_2018,HuangMoXu_WCNC2020,HBayerlein2020_AIforUAV,XLiu20TVT_AIforUAV,CLiu2020TMC_AIforUAV,MChen2019TWC_AIforUAV,JHu020Ar_AIforUAV} and cellular-connected UAVs \cite{YZeng2020Ar_AIforUAV,EFonseca2020Ar_AIforUAV}, when the UAVs serve as BSs and users, respectively, as  elaborated in the following.   {Besides, a comparison of existing works in this field is provided in Table \ref{Table:AI}.}

\begin{table*}[]
\centering
\renewcommand{\arraystretch}{1.6}
{\caption{Comparison of existing works on AI-integration in UAV communication networks.}\label{Table:AI}
{\begin{tabular}{|p{4cm} | p{2cm} |p{9cm}|}
\hline
& \textbf{Reference} & \textbf{Main contributions} \\  
\hline
\multirow{6}{*}
{\textbf{Machine Learning for UAVs}}&
 \cite{Bayerlein_UAV_RL_2018,HuangMoXu_WCNC2020} &Q-learning-based trajectory design for one UAV communicating with multiple ground users\\  
\cline{2-3}
& \cite{HBayerlein2020_AIforUAV} & Joint exploitation of double deep Q-network (DDQN) and environment maps for trajectory control in UAV-enabled data collection \\ 
\cline{2-3}
& \cite{XLiu20TVT_AIforUAV} & Three-step machine learning based approach for optimizing cell partition, 3D deployments, and dynamic movements in multi-UAV-enabled networks\\
\cline{2-3}
& \cite{CLiu2020TMC_AIforUAV} & Decentralized deep RL for multiple UAV-BSs' trajectories design to minimize the AoI  \\
\cline{2-3}
& \cite{YZeng2020Ar_AIforUAV} & DDQN for simultaneous navigation and radio mapping  \\
\cline{2-3}
&\cite{EFonseca2020Ar_AIforUAV} & Deep Q-learning for optimizing the height of UAV to balance between desirable BS's signals and other BSs' inteference \\
\hline
\multirow{5}{*}{\textbf{Computation Offloading}}&
\cite{CaoXuZhang_UAV_MEC_2018} &Joint trajectory and TDMA-based transmission design for one UAV-edge-device to offload computation tasks to multiple BSs for parallel execution\\  
\cline{2-3}
& \cite{SJeong2018TVT_MEC,ZYang2019TWC_MEC,HGuo2020TII_MEC} & Joint communication, computation, and trajectory optimization for UAV-edge-server to maximize the computation performance of ground devices \\ 
\cline{2-3}
& \cite{XHu2019TWC_MEC} & UAV-relay-assisted MEC
 \\
\cline{2-3}
& \cite{FZhou2018JSAC_MEC,YLiu2020IoTJ_MEC} &  UAV-enabled wireless powered MEC \\
\cline{2-3}
& \cite{JZhang2020TVT_MEC,AAsheralieva2019IoTJ_MEC} & Joint trajectory and resource allocation design for multi-UAV-enabled MEC networks\\
\hline
\multirow{4}{*}{\textbf{Edge Machine Learning}}&
\cite{TZeng2020Ar_UAVforAI} &Wireless resource allocation for UAV-enabled federated edge learning\\  
\cline{2-3}
& \cite{YLiu2020Ar_UAVforAI} & Federated edge learning for UAV swarms
 \\
 \cline{2-3}
& \cite{CDong2020Ar_UAV_Service} & UAVs as edge servers for federated learning \\ 
\cline{2-3}
& \cite{JNg2020Ar_UAVforAI} & UAV-enabled federated learning for Internet of vehicles
 \\
\hline
\end{tabular}}}
\end{table*}

\subsubsection{UAVs as BSs}

First, consider the case where one single UAV serves as an aerial BS for UAV-assisted wireless communications and needs to simultaneously serve multiple users on the ground. In this case, the UAV needs to find the optimal flight trajectory maximizing the performance, while taking into account the wireless channel conditions of all users. However, as obstacles (e.g., buildings or trees) are non-uniformly  distributed over space, the UAV may possess LoS and non-LoS (NLoS) links with different ground users at different locations. In practice, as the obstacles' locations may not be known prior to the optimization, the LoS/NLoS property of the wireless channels becomes uncertain, thus making it infeasible to formulate the UAV trajectory design as an explicit optimization problem. Although prior works assumed probabilistic LoS channel models and accordingly used average channel power gains in their problem formulations \cite{zeng2019accessing}, such designs would lead to highly compromised solutions. In contrast, RL is an efficient model-free method, which can learn the environment on-the-fly based on the past experiences, and automatically generate optimized solutions. For instance, the authors in \cite{Bayerlein_UAV_RL_2018} assumed that one UAV communicates with multiple ground users over orthogonal channels, where the UAV aims to maximize the communication sum rate for all users by optimizing the trajectory. By applying Q-learning (a model-free RL technique), the UAV can autonomously learn its optimal trajectory for sum-rate maximization, without explicitly having the environment information. Furthermore, the authors in \cite{HuangMoXu_WCNC2020} studied the sum-rate maximization over a finite duration in a UAV-enabled uplink NOMA system for data collection, where the ground users may move dynamically over time. By assuming  that the UAV only causally knows the users' (moving) locations and CSI, a Q-learning-based RL approach was proposed to enable the UAV to optimize its flying path in an online manner, where expert knowledge of well-established wireless channel models was exploited to initialize the Q-table values.

Besides the model-free design without any LoS/NLoS information in \cite{Bayerlein_UAV_RL_2018,HuangMoXu_WCNC2020}, environment maps \cite{OEsrafilian2019IoTJ_AIforUAV,HBayerlein2020_AIforUAV}, which contain the LoS/NLoS channel information over space, can be further exploited for more efficient trajectory design. For example, the authors in \cite{HBayerlein2020_AIforUAV} considered the UAV-enabled data collection over a finite flight duration subject to obstacle avoidance constraints, for which a new end-to-end RL approach was proposed to efficiently control the UAV trajectory for enhancing the communication performance. For this setup, a double deep Q-network (DDQN) with combined experience replay was trained to learn the UAV control policy. By exploiting a multi-layer map of the environment, the DDQN is applicable  for general scenarios when the system and channel parameters (such as the number of ground users, their locations, and the maximum flight duration) can vary.

Next, when multiple UAVs are deployed in a network for serving multiple users on the ground, it becomes important for the UAVs to jointly design their trajectories and communications to collaboratively enhance the network coverage and throughput performance. For instance, in \cite{XLiu20TVT_AIforUAV}, a three-step machine learning based approach was proposed to maximize the satisfaction of the users, where the cell partition, the 3D deployments of multiple UAVs, and their dynamic movements are sequentially optimized via K-means and two Q-learning-based algorithms, respectively. Furthermore, decentralized deep RL based trajectory designs were developed to ensure coverage and user fairness while maximizing the energy efficiency of UAVs \cite{CLiu2020TMC_AIforUAV}  and to minimize the age-of-information (AoI) when the UAVs execute sensing tasks through cooperative sensing and transmission \cite{JHu020Ar_AIforUAV}. Moreover, the authors in \cite{MChen2019TWC_AIforUAV} considered the scenario of UAV-enabled caching, where a distributed algorithm based on the machine learning framework of liquid state machine (LSM) was developed to optimize the content caching and resource allocation decisions at the UAVs.

\subsubsection{UAVs as Users}

In cellular-connected UAV scenarios, UAVs act as aerial users in cellular networks to perform tasks like packet delivery and aerial inspection. In this scenario, how to maintain the cellular connection while addressing the task is the main concern. Normally, conventional designs optimize the UAV trajectories based on distance-dependent channel models under a fixed path loss exponent. These works, however, overlook the complicated air-ground channel propagation environments, and thus may lead to unexpected communication outage at the cellular BSs in coverage holes where the wireless channels of the ground BSs are blocked by obstacles such as  high buildings. To tackle this challenge, using machine learning based designs for refining the UAV trajectories are emerging to maximize the connectivity probability during a task. For instance, in the case with one single UAV, the authors in \cite{YZeng2020Ar_AIforUAV} proposed a deep RL-based solution approach, namely a dueling double deep Q-network (dueling DDQN), to design the navigation/trajectory for optimally balancing between the UAV mission completion time versus the expected duration of communication outage. More specifically, this design exploits the dual use of the UAV's signal measurements not only for training the DQN, but also for creating a radio map to predict the outage probabilities at different locations, thus leading to an interesting framework termed simultaneous navigation and radio mapping (SNARM) that greatly improves the learning performance.

Besides the signal blockage due to obstacles in the communication path, the co-channel interference among nearby BSs is another issue that may affect the system performance. Recall that the signal path loss for UAV-ground channels is altitude-dependent, i.e., when the UAV stays at a higher altitude, the LoS probability of the UAV-ground channels increases. Exploiting such properties,  deep Q-learning  was used in \cite{EFonseca2020Ar_AIforUAV}  for optimizing the height of the UAV for maximization of the spectral efficiency, via properly balancing between the desirable signal and harmful interference.

\begin{figure*}[t]
\centering
 \setlength{\abovecaptionskip}{-0mm}
\setlength{\belowcaptionskip}{-0mm}
    \includegraphics[width=6in]{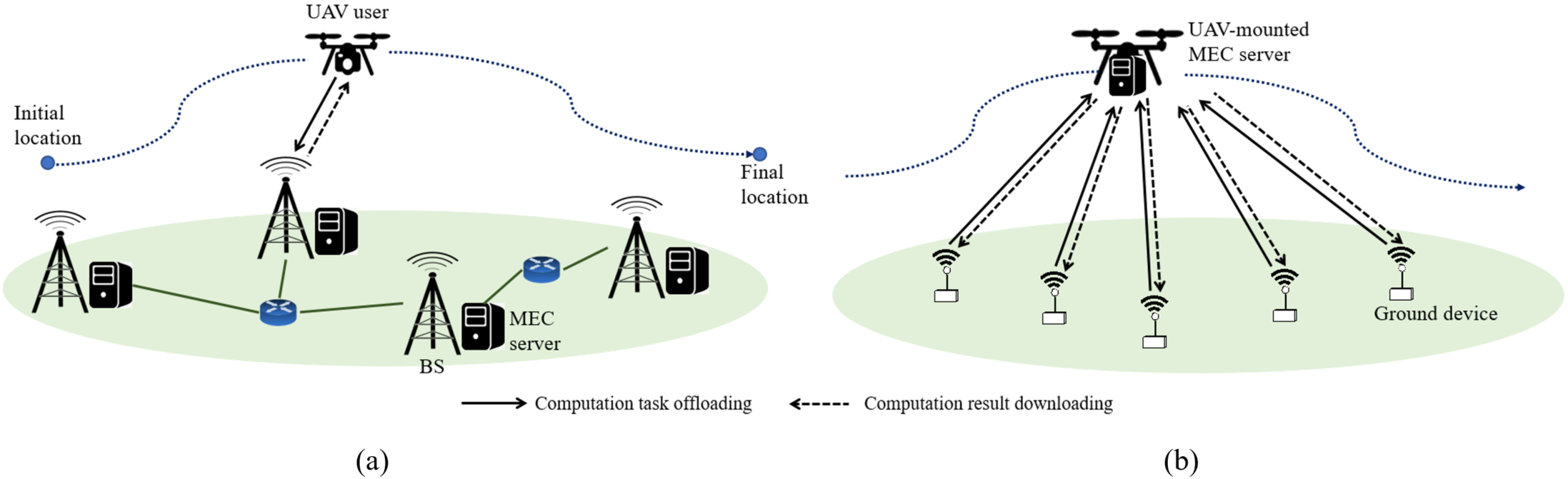}
\caption{Computation offloading with UAV: (a) cellular-connected UAV; (b) UAV-assisted MEC system. } \label{fig:UAVOffloading}
\end{figure*}

\subsection{Computation Offloading with UAVs for AI Tasks}

To enable edge intelligence applications with both UAVs and AI integrated, a large volume of data are to be generated at distributed edge devices (including both UAVs and conventional smart sensors and smart phones), which need to be properly processed via sophisticated AI training and inference algorithms. However, the implementation of these AI tasks is generally data- and computation-intensive, which cannot be handled locally by wireless devices themselves. As such, computation task offloading is an appealing solution to handle AI tasks, which allows the UAV edge servers to offload their AI tasks to MEC servers with high computation capabilities, and then download the computation results after MEC execution. Fig.~\ref{fig:UAVOffloading}(a) shows the scenario when a UAV acts as an aerial user or edge device in cellular networks, which has certain computation tasks to be executed via offloading to BSs on the ground. Besides edge devices, UAVs may also carry MEC servers to support the AI implementation of on-ground devices. As shown in Fig.~\ref{fig:UAVOffloading}(b), the UAV can help the widely distributed devices on the ground to execute their computation-intensive AI tasks, especially in emergency situations. Under both scenarios, the joint UAV trajectory and communication/computation design is crucial. For facilitate such joint design, in this line of research the AI tasks are normally modeled as general computation tasks with certain data and computation requirements, as  detailed in the following.

\subsubsection{UAVs as Edge Devices}

First, as shown in Fig.~\ref{fig:UAVOffloading}(a), assume that a UAV user needs to fly from an initial to a final location (e.g., for packet delivery) and  has certain computation tasks to be executed during the mission (e.g., processing of sensed data in real time for autonomous flight). During the flight, the UAV needs to offload its computation tasks to ground BSs/MEC servers, and download the computation results from the corresponding BSs after their remote execution. In this case, the UAV's trajectory design becomes a more complicated problem due to the distributed computation constraints at the BSs. For instance, a UAV may prefer to partition the computation tasks and offload them to different BSs to exploit their cooperative computation gain; and accordingly, the trajectory must be carefully design in order to meet the offloading-execution-downloading requirements of different BSs. It should be further noted that the UAV can only download the computation results from the same BS to which the tasks were offloaded, thus resulting in the coupling between (uplink) offloading and (downlink) downloading. This makes the trajectory design even more difficult, as the UAV may need to fly back and forth during offloading and downloading.

\begin{figure}
\centering
 \setlength{\abovecaptionskip}{-1mm}
\setlength{\belowcaptionskip}{-1mm}
    \includegraphics[width=3.5in]{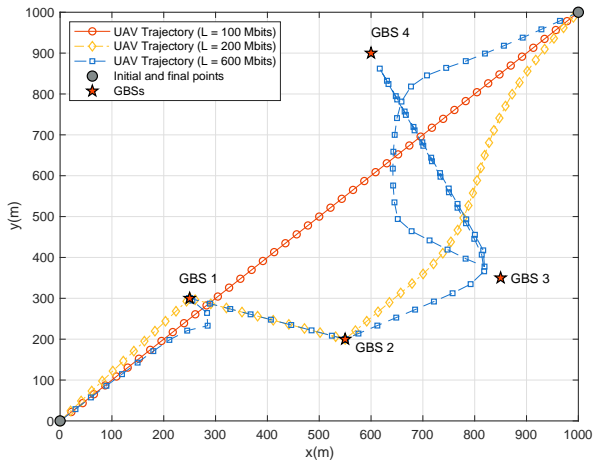}
\caption{Obtained UAV trajectory for computation offloading for different computation loads. } \label{fig:Mobility_K4}
\end{figure}

There have been various attempts to address the aforementioned problems. For example, the authors in \cite{CaoXuZhang_UAV_MEC_2018} proposed a TDMA-based protocol for computation offloading, where the UAVs can offload their tasks to different BSs for parallel execution in orthogonal time slots, in order to fully exploit the computation resources available at the distributed BSs. In particular, the UAV's mission completion time is minimized by jointly optimizing the UAV trajectory and the computation offloading scheduling, subject to the UAV's flight constraints, and the ground BSs' (GBSs') computation capacity constraints. For illustration, Fig.~\ref{fig:Mobility_K4} shows the optimized UAV trajectories for computation offloading for different computation loads (measured in computation task input bits $L$ \cite{CaoXuZhang_UAV_MEC_2018}).  There are four GBSs and the UAV's initial and final coordinates  are  (0,0) and (1000 m, 1000 m), respectively. It is observed that when the computation load is small (i.e., $L =$ 100 Mbits), the UAV flies in a straight line from the initial to the final location. When $L$ increases to 200 Mbits, the flight trajectory deviates from the straight line for more efficient offloading. For $L=$ 600 Mbits, the UAV flies back and forth between different GBSs in order to exploit multiple GBSs' distributed computation resources more efficiently via time sharing. This phenomenon is significantly different from the UAV trajectory design for communications, where the UAV may successively visit different ground nodes, instead of flying back and forth for computation offloading because of the inherent distributed computation capacity constraints.

\subsubsection{UAVs as Edge Servers}

\begin{figure*}
\centering
 \setlength{\abovecaptionskip}{-1mm}
\setlength{\belowcaptionskip}{-1mm}
    \includegraphics[width=6in]{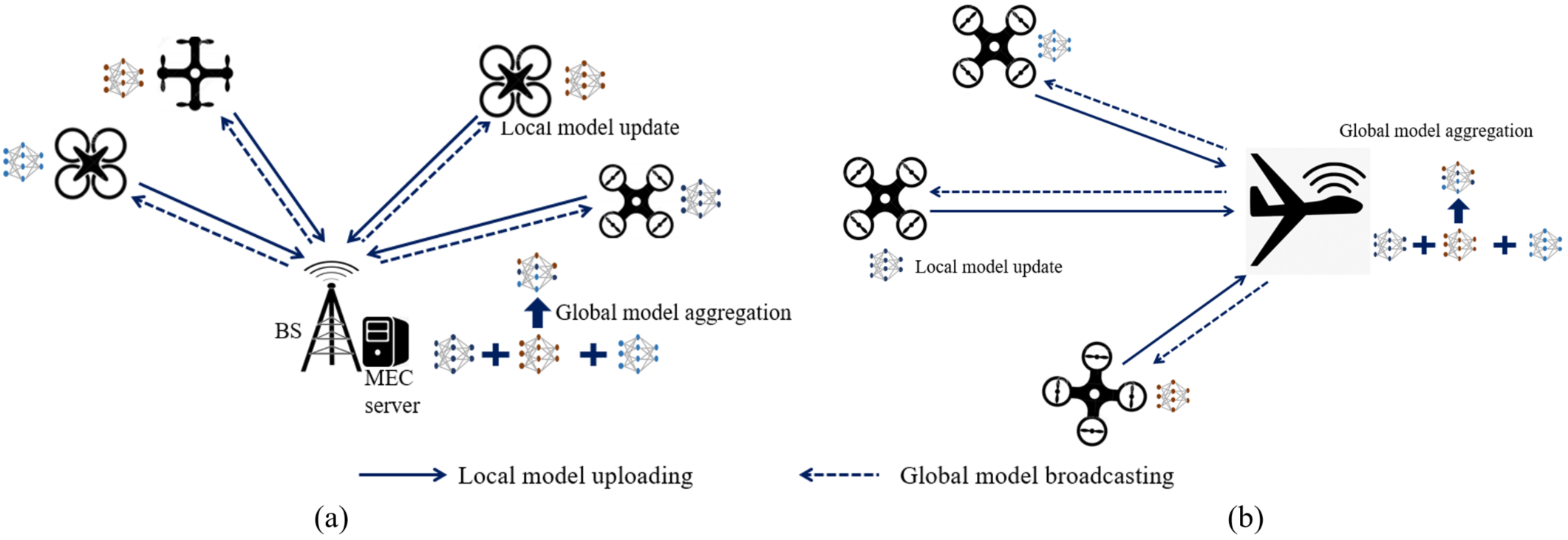}
\caption{Federated edge learning with UAVs: (a) the case with coordination of ground BS; (b) the case with UAV peers. } \label{fig:UAVLearning}
\end{figure*}

In another line of research, the UAVs can be utilized as MEC servers in the sky to  remotely execute the computation tasks of ground devices, as shown in Fig.~\ref{fig:UAVOffloading}(b), which may correspond to scenarios when the ground infrastructure is damaged or not available. As the UAV's limited communication and computation resources are shared among multiple ground devices, the UAV should carefully design the joint communication and computation scheduling, together with the trajectory design, to maximize the computation performance of ground devices in a fair and efficient manner \cite{SJeong2018TVT_MEC,ZYang2019TWC_MEC,HGuo2020TII_MEC}.

Furthermore, the single-UAV-assisted computation offloading in Fig.~\ref{fig:UAVOffloading}(b) has been extended to various other setups. For instance, the authors in \cite{XHu2019TWC_MEC} presented a UAV-relay-assisted MEC system by integrating UAVs into the cooperative MEC design \cite{Cao2019IoTJ}, where the UAV not only acts as the MEC server for remote task execution, but also serves as a relay to help ground devices offload tasks to another GBS. In this system, the UAV exploits the joint communication and computation cooperation for computation performance enhancement. Moreover , the authors in \cite{FZhou2018JSAC_MEC,YLiu2020IoTJ_MEC} proposed UAV-enabled MEC wireless-powered systems by integrating the UAV with wireless powered MEC \cite{WangXu_WirelessPoweredMEC_2018}, where the UAV serves as  energy transmitter, communication transceiver, and MEC server. In this system, the UAV trajectory is optimized jointly with the energy/communication/computation resource allocations for  maximization of  the computation performance of the ground devices, subject to a new type of wireless energy harvesting constraint as the energy consumption for communication and computation at each device cannot exceed the energy harvested wirelessly from the UAV.  In addition, security concerns during offloading from ground devices to the UAV were addressed in \cite{TBai2019TVT_MEC}, where  physical layer security  is employed to combat against potential eavesdropping from ground attackers.

There have also been a handful of works investigating UAV-assisted computation offloading to multiple UAV-mounted MEC servers \cite{JZhang2020TVT_MEC,AAsheralieva2019IoTJ_MEC}. In this case, the association between the ground devices and the UAVs is an important new issue to be considered, together with UAV trajectory and resource allocation design \cite{JZhang2020TVT_MEC}. Different from conventional designs focusing on communication only, the user association in the MEC context needs to properly balance both the communication and computation loads, thus making it more difficult. Furthermore, the self-interest is another issue for multi-UAV networks, especially when the UAVs may belong to different service providers \cite{AAsheralieva2019IoTJ_MEC}. To address this issue, coalition formation  for UAVs can be investigated by using game theory and RL.

Moreover, researchers also investigated UAV swarm applications with stringent computation requirements. Towards this end, one efficient solution is to employ high-altitude UAVs with larger size as edge servers to support the swarm UAVs' computation \cite{QZhang2020TVT_MEC}, while another solution is to use the edge and cloud infrastructures on the ground to support the swarm UAVs' extensive computation tasks  \cite{WChen2019Networks_EdgeComputing}.

\subsection{Distributed Edge Machine Learning with UAVs}

Besides computation offloading towards a central MEC server,  edge devices such as UAVs can also collaboratively perform AI tasks using their locally distributed data and computation capabilities. This technique is generally referred to as distributed edge learning. Among other approaches, federated edge learning, {initially developed by Google \cite{FederatedLearning2016A},} is particularly important due to its advantages in preserving  data security and privacy \cite{Guangxu_edge_lerning2020}, {where a group of edge devices (UAVs of interest) jointly use their distributed data to train common AI or machine learning models without data sharing.} As illustrated in Fig.~\ref{fig:UAVLearning}(a) and Fig.~\ref{fig:UAVLearning}(b), the edge server can either be a BS on the ground (e.g.,  as with a cellular-connected UAV) or another UAV in the sky (e.g., as in a UAV swarm). Federated edge learning is implemented in an iterative manner: in each iteration, the UAVs first update their local AI models, and then aggregate at the edge server to update global AI models. The above process requires the UAVs and edge server to frequently exchange their AI model parameters, and iterate between communication and computation operations. Because of this fact and the UAVs' 3D mobility, how to jointly optimize their multiple UAVs' trajectories together with the communication and computation scheduling over time is the key issue to be tackled.

While federated edge learning is currently a very hot topic in both the wireless communications and machine learning societies, the research on UAV-enabled edge learning is still in its infancy. For instance, {\cite{AIIntegration2020C} presented various applications of federated learning in UAV wireless communication networks.} Wireless resource allocation  for enabling efficient UAV-enabled federated edge learning was studied in  \cite{TZeng2020Ar_UAVforAI}. Federated learning for UAV swarms was investigated in \cite{YLiu2020Ar_UAVforAI}. Furthermore, using UAVs as edge servers for federated learning was proposed in \cite{CDong2020Ar_UAV_Service}, and the support of Internet of vehicles was considered in  \cite{JNg2020Ar_UAVforAI}. {Despite this research progress, the fundamental performance limits of federated learning with mobile UAV nodes are still a largely uncharted area.}

\subsection{Future Research}

\subsubsection{Cloud-Edge-Device Collaboration with UAVs}

As UAVs may perform as both aerial users and service (communication/computation/AI) providers, future networks need to be aerial-ground integrated, where computation resources are distributed everywhere across heterogeneous aerial and ground nodes with distinct communication and computation capabilities (see, e.g., \cite{WZhang2020Access_UAVMECSurvey}). How to match the time- and spatial-varying communication/computation/AI demands with distributed communication/computation/data supplies in such highly dynamic 3D networks is a challenging task.

\subsubsection{Over-the-Air Computation From Sky}

Over-the-air computation (AirComp) is an emerging technique to enable functional computation over wireless channels by exploiting wireless signal superposition property of multiple access channels \cite{Cao_AirComp_2020,Zhu_MIMO_AirComp_2019,Nazer2007,Cao2020Ar_MultiCellAirComp,ZhuXuHuang_AirComp_2020}. AirComp leads to a fully-integrated-communication-computation design, and thus can have abundant applications in distributed sensing, computation, federated learning \cite{Zhu_FL_AirComp_2020,Yang_FL_AirComp_2020}, and UAV swarms. When UAVs are involved in AirComp as either edge devices or servers, how to exploit their mobility for optimizing the AirComp performance is an interesting problem that is uncharted yet.

\section{Conclusions}
In this paper, we provided a comprehensive overview on the integration of UAVs into 5G-and-beyond  networks, which is expected to drive  UAV verticals and cellular networks to a win-win situation by jointly exploiting their high altitude and  flexible 3D movement  together with the most advanced wireless technologies.   In addition to  reviewing the state-of-the-art research from both the academic and industrial  viewpoints, we  focused on the most promising  solutions of addressing the challenges in communication, sensing, and AI, and also exploring new opportunities in system design and implementation of UAV cellular networks. Promising directions and open problems in 3D wireless networks with UAVs were also highlighted for future research. As the research in this frontier paradigm is new and remains largely unexplored, it is hoped that this paper will serve as a timely and useful resource for researchers to push forward.



\bibliographystyle{IEEEtran}
\bibliography{IEEEabrv,mybib}

\begin{IEEEbiography}[{\includegraphics[width=1in,height=1.25in,clip,keepaspectratio]{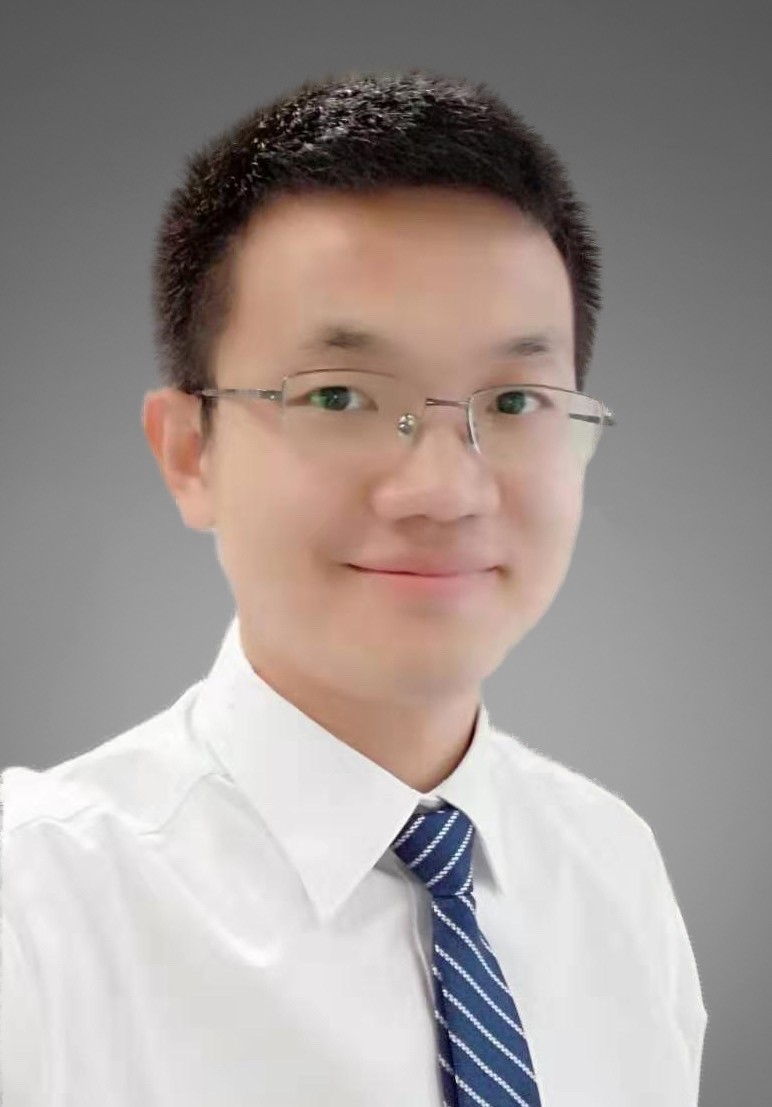}}]
{Qingqing Wu} (S'13-M'16) received the B.Eng. and the Ph.D. degrees in Electronic Engineering from South China University of Technology and Shanghai Jiao Tong University (SJTU) in 2012 and 2016, respectively. He is currently an assistant professor with the State key laboratory of Internet of Things for Smart City, University of Macau. From 2016 to 2020, he was a Research Fellow in the Department of Electrical and Computer Engineering at National University of Singapore. His current research interest includes intelligent reflecting surface (IRS), unmanned aerial vehicle (UAV) communications, and MIMO transceiver design. He was listed as a World's Top 2\% Scientist by Stanford University in 2020.

He was the recipient of the IEEE Communications Society Young Author Best Paper Award in 2021, the Outstanding Ph.D. Thesis Award of China Institute of Communications in 2017, the Outstanding Ph.D. Thesis Funding in SJTU in 2016, and the IEEE WCSP Best Paper Award in 2015. He was the Exemplary Editor of IEEE Communications Letters in 2019 and the Exemplary Reviewer of several IEEE journals. He serves as an Associate Editor for IEEE Communications Letters, IEEE Wireless Communications Letters, IEEE Open Journal of Communications Society (OJ-COMS), and IEEE Open Journal of Vehicular Technology (OJVT). He is the Lead Guest Editor for IEEE Journal on Selected Areas in Communications on UAV Communications in 5G and Beyond Networks, and the Guest Editor for IEEE OJVT on 6G Intelligent Communications and IEEE OJ-COMS on Reconfigurable Intelligent Surface-Based Communications for 6G Wireless Networks. He is the workshop co-chair for IEEE ICC 2019-2021 workshop on Integrating UAVs into 5G and Beyond, and the workshop co-chair for IEEE GLOBECOM 2020 and ICC 2021 workshop on Reconfigurable Intelligent Surfaces for Wireless Communication for Beyond 5G. He serves as the Workshops and Symposia Officer of Reconfigurable Intelligent Surfaces Emerging Technology Initiative and Research Blog Officer of Aerial Communications Emerging Technology Initiative. He is the IEEE ComSoc Young Professional Chair in APB.

\end{IEEEbiography}

\begin{IEEEbiography}[{\includegraphics[width=1in,height=1.25in,clip,keepaspectratio]{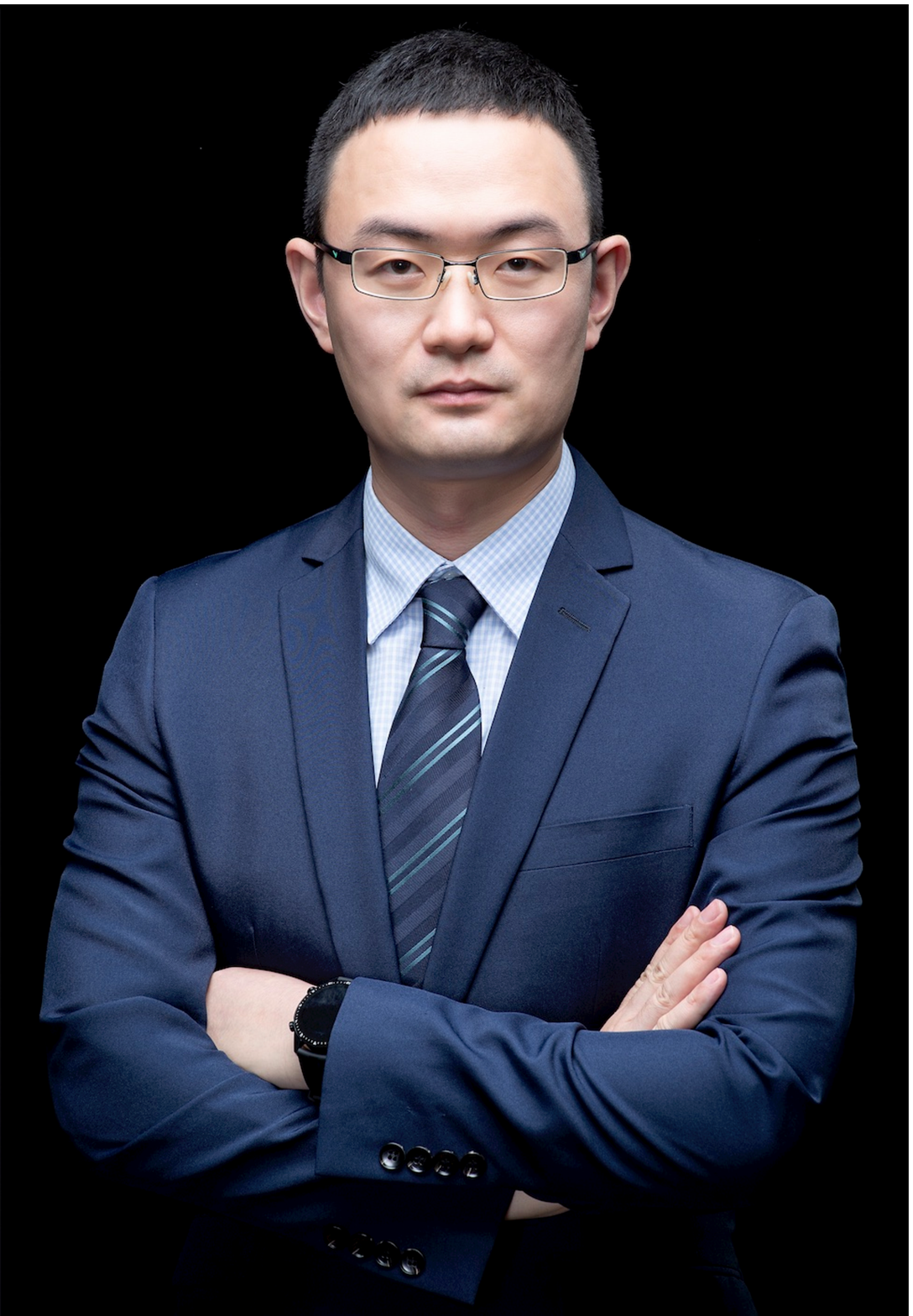}}]
{Jie Xu} (S'12-M'13) received the B.E. and Ph.D. degrees from the University of Science and Technology of China in 2007 and 2012, respectively. From 2012 to 2014, he was a Research Fellow with the Department of Electrical and Computer Engineering, National University of Singapore. From 2015 to 2016, he was a Post-Doctoral Research Fellow with the Engineering Systems and Design Pillar, Singapore University of Technology and Design. From 2016 to 2019, he was a Professor with the School of Information Engineering, Guangdong University of Technology, China. He is currently an Associate Professor with the School of Science and Engineering, The Chinese University of Hong Kong, Shenzhen, China. His research interests include wireless communications, wireless information and power transfer, UAV communications, edge computing and intelligence, and integrated sensing and communication (ISAC). He was a recipient of the 2017 IEEE Signal Processing Society Young Author Best Paper Award, the IEEE/CIC ICCC 2019 Best Paper Award, the 2019 IEEE Communications Society Asia-Pacific Outstanding Young Researcher Award, and the 2019 Wireless Communications Technical Committee Outstanding Young Researcher Award. He is the Symposium Co-Chair of the IEEE GLOBECOM 2019 Wireless Communications Symposium, the workshop co-chair of several IEEE ICC and GLOBECOM workshops, the Tutorial Co-Chair of the IEEE/CIC ICCC 2019, and the Founding Chair of the IEEE WTC Special Interest Group (SIG) on ISAC. He served or is serving as an Editor of the IEEE TRANSACTIONS ON COMMUNICATIONS, IEEE WIRELESS COMMUNICATIONS LETTERS, and Journal of Communications and Information Networks, an Associate Editor of IEEE ACCESS, and a Guest Editor of the IEEE WIRELESS COMMUNICATIONS, IEEE JOURNAL ON SELECTED AREAS IN COMMUNICATIONS, and Science China Information Sciences.
\end{IEEEbiography}

\begin{IEEEbiography}[{\includegraphics[width=1in,height=1.25in,clip,keepaspectratio]{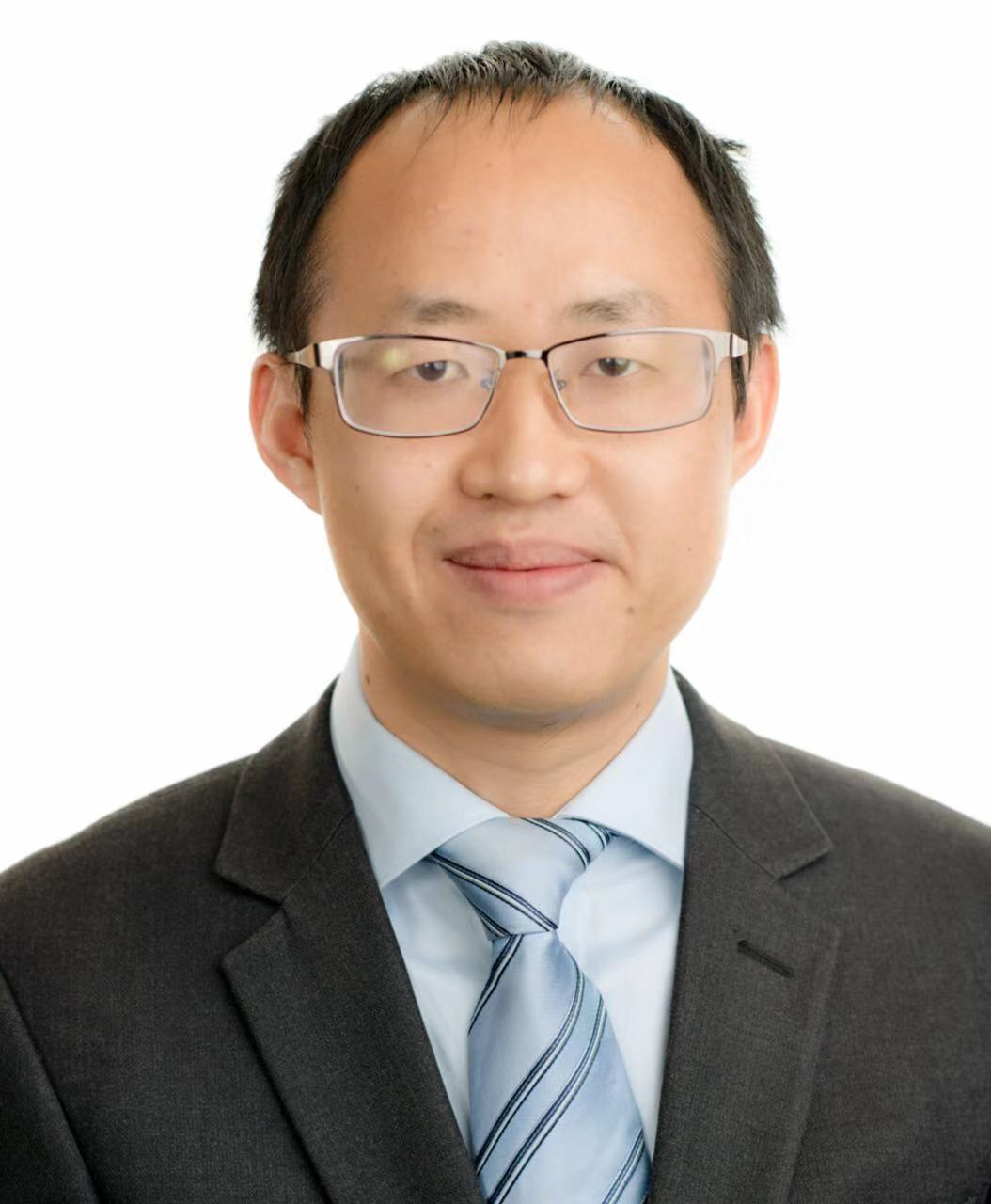}}]
{Yong Zeng} (S'12-M'14) is with the National Mobile Communications Research Laboratory, Southeast University, China, and also with the Purple Mountain Laboratories, Nanjing, China. He received the Bachelor of Engineering (First-Class Honours) and Ph.D. degrees from Nanyang Technological University, Singapore, in 2009 and 2014, respectively. From 2013 to 2018, he was a Research Fellow and Senior Research Fellow at the Department of Electrical and Computer Engineering, National University of Singapore. From 2018 to 2019, he was a Lecturer at the School of Electrical and Information Engineering, the University of Sydney, Australia.

Dr. Zeng was listed as 2020 and 2019 Highly Cited Researcher by Clarivate Analytics. He is the recipient of the Australia Research Council (ARC) Discovery Early Career Researcher Award (DECRA), 2020 IEEE Marconi Prize Paper Award in Wireless Communications, 2018 IEEE Communications Society Asia-Pacific Outstanding Young Researcher Award, 2020 \& 2017 IEEE Communications Society Heinrich Hertz Prize Paper Award. He serves as an Associated Editor for IEEE Communications Letters and IEEE Open Journal of Vehicular Technology, Leading Guest Editor for IEEE Wireless Communications on ``Integrating UAVs into 5G and Beyond'' and China Communications on ``Network-Connected UAV Communications''. He is the workshop co-chair for ICC 2018-2021 workshop on UAV communications, the tutorial speaker for Globecom 2018/2019 and ICC 2019 tutorials on UAV communications.
\end{IEEEbiography}

\begin{IEEEbiography}[{\includegraphics[width=0.95in,height=1.4in]{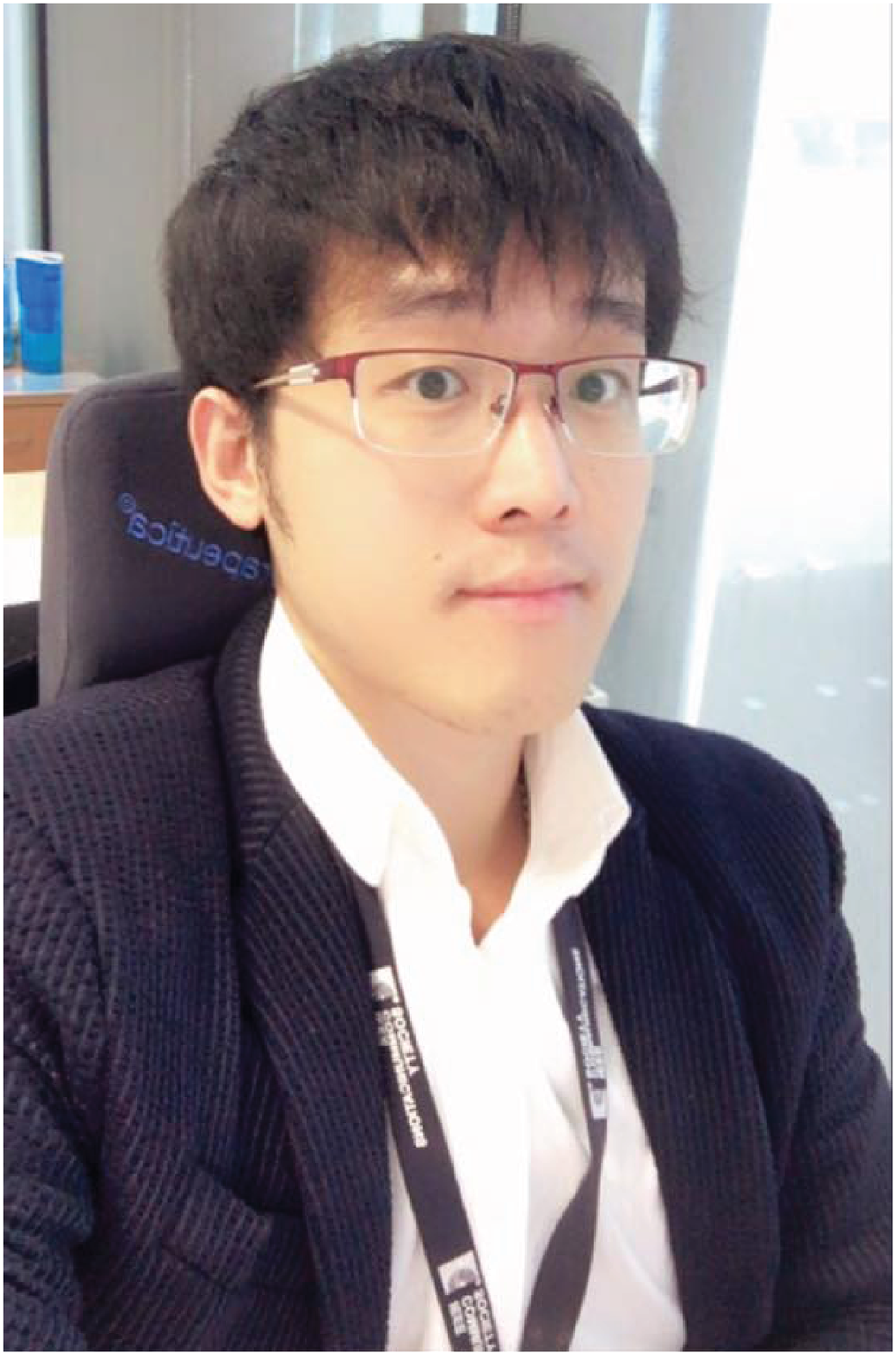}}]{Derrick
Wing Kwan Ng} (S'06-M'12-SM'17-F'21)   received the bachelor degree with first-class honors and the Master of Philosophy (M.Phil.) degree in electronic engineering from the Hong Kong University of Science and Technology (HKUST) in 2006 and 2008, respectively. He received his Ph.D. degree from the University of British Columbia (UBC) in 2012. He was a senior postdoctoral fellow at the Institute for Digital Communications, Friedrich-Alexander-University Erlangen-N\"urnberg (FAU), Germany. He is now working as a Senior Lecturer and a Scientia Fellow at the University of New South Wales, Sydney, Australia.  His research interests include convex and non-convex optimization, physical layer security, IRS-assisted communication, UAV-assisted communication, wireless information and power transfer, and green (energy-efficient) wireless communications.

Dr. Ng received the Australian Research Council (ARC) Discovery Early Career Researcher Award 2017,   the Best Paper Awards at the WCSP 2020,  IEEE TCGCC Best Journal Paper Award 2018, INISCOM 2018, IEEE International Conference on Communications (ICC) 2018,  IEEE International Conference on Computing, Networking and Communications (ICNC) 2016,  IEEE Wireless Communications and Networking Conference (WCNC) 2012, the IEEE Global Telecommunication Conference (Globecom) 2011, and the IEEE Third International Conference on Communications and Networking in China 2008.  He has been serving as an editorial assistant to the Editor-in-Chief of the IEEE Transactions on Communications from Jan. 2012 to Dec. 2019. He is now serving as an editor for the IEEE Transactions on Communications,  the IEEE Transactions on Wireless Communications, and an area editor for the IEEE Open Journal of the Communications Society. Also, he has been listed as a Highly Cited Researcher by Clarivate Analytics since 2018.
\end{IEEEbiography}

\begin{IEEEbiography}[{\includegraphics[width=1in,height=1.25in,clip,keepaspectratio]{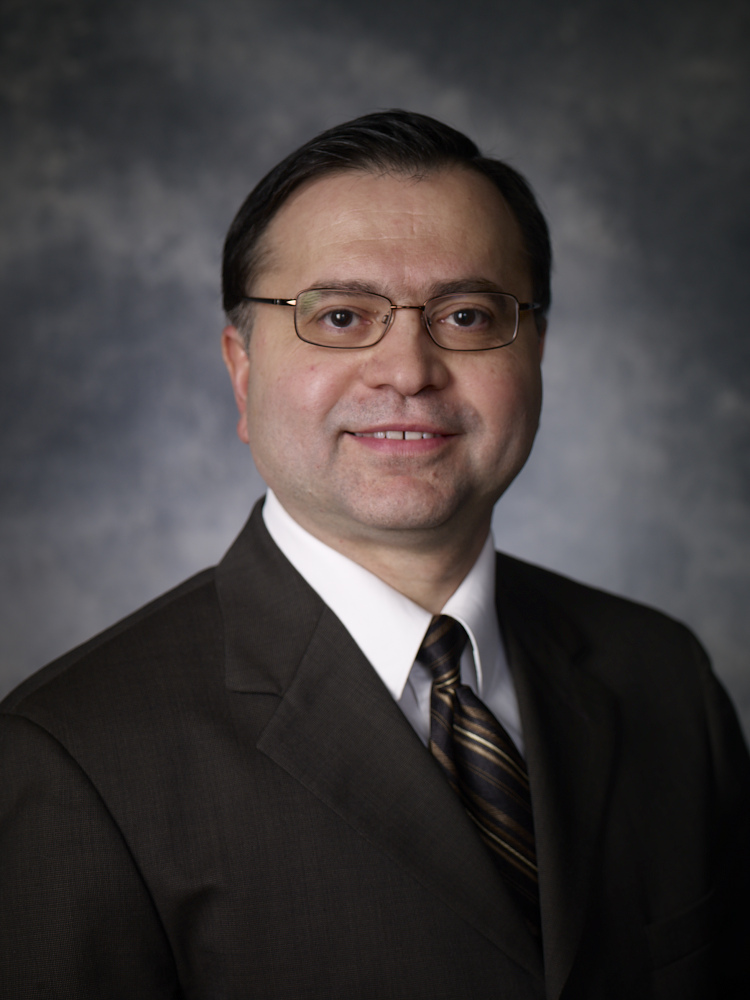}}]
{Naofal Al-Dhahir} is Erik Jonsson Distinguished Professor \& ECE Dept. Associate Head at UT-Dallas. He earned his PhD degree from Stanford University and was a principal member of technical staff at GE Research Center and AT\&T Shannon Laboratory from 1994 to 2003.  He is co-inventor of 43 issued patents, co-author of about 470 papers and co-recipient of 4 IEEE best paper awards. He is an IEEE Fellow, received 2019 IEEE SPCC technical recognition award and 2021 Qualcomm faculty award. He served as Editor-in-Chief of IEEE Transactions on Communications from Jan. 2016 to Dec. 2019.  He is a Fellow of the National Academy of Inventors.
\end{IEEEbiography}

\begin{IEEEbiography}[{\includegraphics[width=1in,height=1.25in,clip,keepaspectratio]{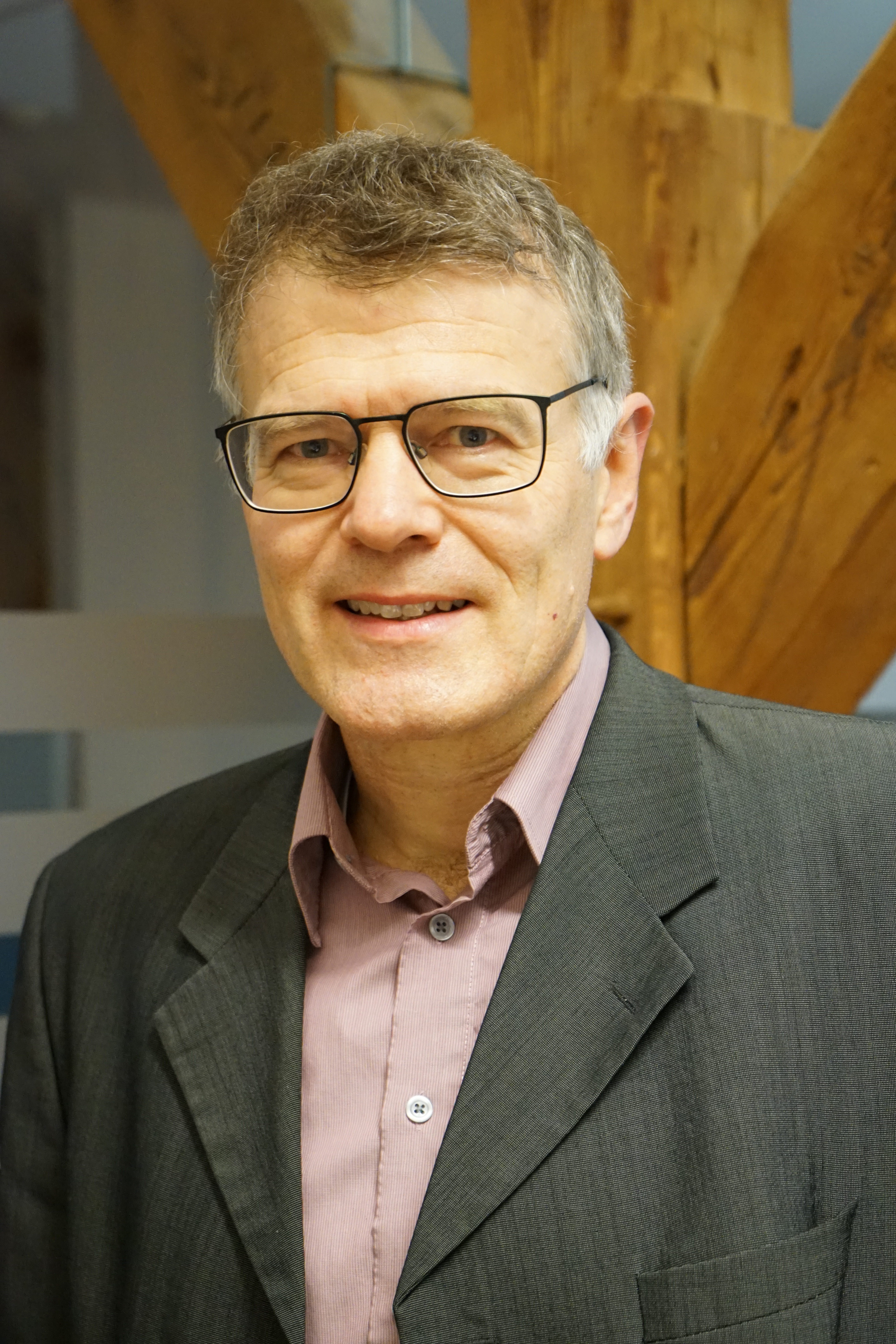}}]
{Robert Schober} (S'98-M'01-SM'08-F'10) received the Diplom (Univ.) and the Ph.D. degrees in electrical engineering from Friedrich-Alexander University of Erlangen-Nuremberg (FAU), Germany, in 1997 and 2000, respectively. From 2002 to 2011, he was a Professor and Canada Research Chair at the University of British Columbia (UBC), Vancouver, Canada. Since January 2012 he is an Alexander von Humboldt Professor and the Chair for Digital Communication at FAU. His research interests fall into the broad areas of Communication Theory, Wireless Communications, and Statistical Signal Processing.

Robert received several awards for his work including the 2002 Heinz Maier¬ Leibnitz Award of the German Science Foundation (DFG), the 2004 Innovations Award of the Vodafone Foundation for Research in Mobile Communications, a 2006 UBC Killam Research Prize, a 2007 Wilhelm Friedrich Bessel Research Award of the Alexander von Humboldt Foundation, the 2008 Charles McDowell Award for Excellence in Research from UBC, a 2011 Alexander von Humboldt Professorship, a 2012 NSERC E.W.R. Stacie Fellowship, and a 2017 Wireless Communications Recognition Award by the IEEE Wireless Communications Technical Committee. Since 2017, he has been listed as a Highly Cited Researcher by the Web of Science. Robert is a Fellow of the Canadian Academy of Engineering, a Fellow of the Engineering Institute of Canada, and a Member of the German National Academy of Science and Engineering. From 2012 to 2015, he served as Editor-in-Chief of the IEEE Transactions on Communications. Currently, he serves as Member of the
Editorial Board of the Proceedings of the IEEE and as VP Publications for the
IEEE Communication Society (ComSoc).

\end{IEEEbiography}

\begin{IEEEbiography}[{\includegraphics[width=1in,height=1.25in,clip,keepaspectratio]{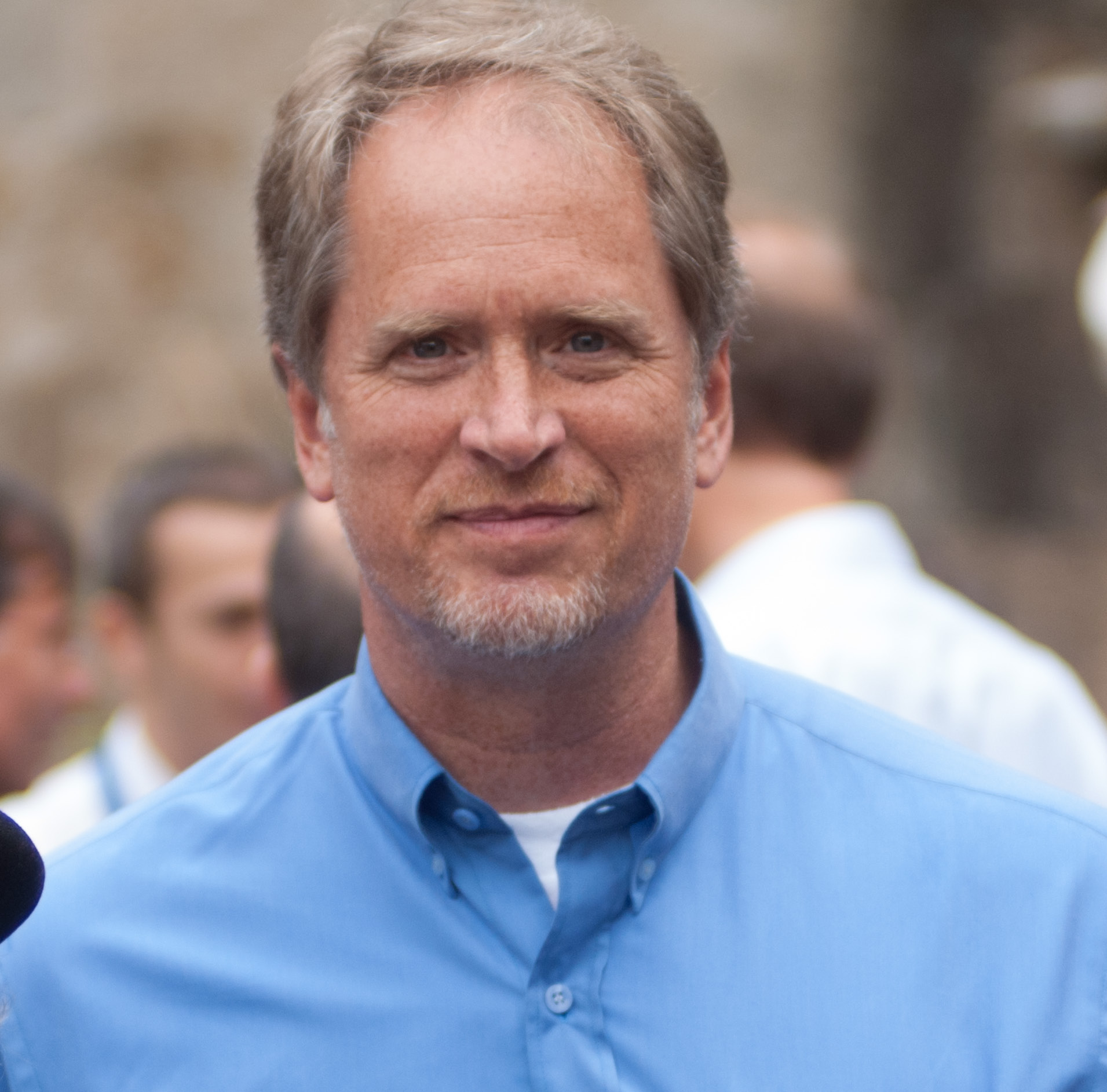}}]
{A. Lee Swindlehurst} received the B.S. (1985) and M.S. (1986) degrees in Electrical Engineering from Brigham Young University (BYU), and the PhD (1991) degree in Electrical Engineering from Stanford University. He was with the Department of Electrical and Computer Engineering at BYU from 1990-2007, where he served as Department Chair from 2003-06.  During 1996-97, he held a joint appointment as a visiting scholar at Uppsala University and the Royal Institute of Technology in Sweden. From 2006-07, he was on leave working as Vice President of Research for ArrayComm LLC in San Jose, California. Since 2007 he has been a Professor in the Electrical Engineering and Computer Science Department at the University of California Irvine, where he served as Associate Dean for Research and Graduate Studies in the Samueli School of Engineering from 2013-16. During 2014-17 he was also a Hans Fischer Senior Fellow in the Institute for Advanced Studies at the Technical University of Munich. In 2016, he was elected as a Foreign Member of the Royal Swedish Academy of Engineering Sciences (IVA). His research focuses on array signal processing for radar, wireless communications, and biomedical applications, and he has over 300 publications in these areas. Dr. Swindlehurst is a Fellow of the IEEE and was the inaugural Editor-in-Chief of the IEEE Journal of Selected Topics in Signal Processing. He received the 2000 IEEE W. R. G. Baker Prize Paper Award, the 2006 IEEE Communications Society Stephen O. Rice Prize in the Field of Communication Theory, the 2006 and 2010 IEEE Signal Processing Society’s Best Paper Awards, and the 2017 IEEE Signal Processing Society Donald G. Fink Overview Paper Award.
\end{IEEEbiography}

\end{document}